\documentclass[twocolumn]{aastex631}
\usepackage{upgreek}
\usepackage{amsmath, amstext}
\usepackage{empheq}
\usepackage{comment}
\usepackage{mathrsfs}
\usepackage{textcomp}
\usepackage{enumitem}   
\usepackage{gensymb}
\usepackage{hyperref}
\usepackage{graphicx}
\usepackage[caption=false]{subfig}
\usepackage{multirow}
\usepackage{longtable}
\usepackage{needspace}
\usepackage{tabularx}
\usepackage[flushleft]{threeparttable}
\usepackage{booktabs} % To thicken table lines
\usepackage{CJK}
\hypersetup{colorlinks, linkcolor={blue}, citecolor={blue}, urlcolor={blue}}

\definecolor{DarkOrange}{RGB}{204, 85, 0}
\definecolor{LincolnGreen}{RGB}{17, 102, 0}

\usepackage{xspace}

\newcommand\nustar{\textit{NuSTAR}\xspace}
\newcommand\nicer{\textit{NICER}\xspace}
\newcommand\chandra{\textit{Chandra}\xspace}
\newcommand\swift{\textit{Swift}\xspace}
\newcommand\xmm{\textit{XMM-Newton}\xspace}

\newcommand\cps{$\rm count\,s^{-1}$\xspace}

\newcommand\target{AT2022cmc\xspace}

\def \caltech {{Cahill Center for Astrophysics, California Institute of Technology, MC 249-17, 1200 E California Boulevard, Pasadena, CA 91125, USA}}

\begin{document}
\pagenumbering{arabic}
%\setmathfont{Cambria Math}

\title{The On-axis Jetted Tidal Disruption Event AT2022cmc:\\
X-ray Observations and Broadband Spectral Modeling
}

%\correspondingauthor{Yuhan Yao}

\author[0000-0001-6747-8509]{Yuhan Yao}\email{yuhanyao@berkeley.edu}
\affiliation{\caltech}
\affiliation{Miller Institute for Basic Research in Science, 468 Donner Lab, Berkeley, CA 94720, USA}
\affiliation{Department of Astronomy, University of California, Berkeley, CA 94720, USA}

\author[0000-0002-1568-7461]{Wenbin Lu}
\affiliation{Department of Astronomy, University of California, Berkeley, CA 94720, USA}
\affiliation{Theoretical Astrophysics Center, University of California, Berkeley, CA 94720, USA}

\author[0000-0003-2992-8024]{Fiona Harrison}
\affiliation{\caltech}

\author[0000-0001-5390-8563]{S. R. Kulkarni}
\affiliation{\caltech}

\author[0000-0003-3703-5154]{Suvi Gezari}
\affiliation{Space Telescope Science Institute, 3700 San Martin Drive, Baltimore, MD 21218, USA}
\affiliation{Department of Physics and Astronomy, Johns Hopkins University, Baltimore, MD 21218}

\author[0000-0002-5063-0751]{Muryel Guolo}
\affiliation{Department of Physics and Astronomy, Johns Hopkins University, Baltimore, MD 21218}

\author[0000-0003-1673-970X]{S.~Bradley Cenko}
\affiliation{Astrophysics Science Division, NASA Goddard Space Flight Center, Greenbelt, MD 20771, USA}
\affiliation{Joint Space-Science Institute, University of Maryland, College Park, MD 20742, USA}

\author[0000-0002-9017-3567]{Anna Y. Q.~Ho}
\affiliation{Department of Astronomy, Cornell University, Ithaca, NY 14853, USA}

\begin{abstract}
AT2022cmc was recently reported as the first on-axis jetted tidal disruption event (TDE) discovered in the last decade, and the fourth on-axis jetted TDE candidate known so far. In this work, we present \nustar hard X-ray (3--30\,keV) observations of AT2022cmc, as well as soft X-ray (0.3--6\,keV) observations obtained by \nicer, \swift, and \xmm. Our analysis reveals that the broadband X-ray spectra can be well described by a broken power-law with $f_\nu \propto \nu^{-0.5}$ ($f_\nu \propto \nu^{-1}$) below (above) the rest-frame break energy of $E_{\rm bk}\sim 10$\,keV at observer-frame $t_{\rm obs}=7.8$ and 17.6\,days since discovery. At $t_{\rm obs} = 36.2$\,days, the X-ray spectrum is consistent with either a single power-law  or a broken power-law. By modeling the spectral energy distribution from radio to hard X-ray across the three \nustar observing epochs, we find that the sub-millimeter/radio emission originates from external shocks at large distances $\gtrsim\! 10^{17}$\,cm from the black hole, the UV/optical light comes from a thermal envelope with radius $\sim\!10^{15}$\,cm, and the X-ray emission is consistent with synchrotron radiation powered by energy dissipation at intermediate radii within the (likely magnetically dominated) jet. We constrain the bulk Lorentz factor of the jet to be of the order 10--100. Our interpretation differs from the model proposed by \citet{Pasham2023} where both the radio and X-rays come from the same emitting zone in a matter-dominated jet. Our model for the jet X-ray emission has broad implications on the nature of relativistic jets in other sources such as gamma-ray bursts.
\end{abstract}

\vspace{1em}

\section{Introduction}
An unlucky star coming too close to a massive black hole (BH) gets disrupted by the tidal forces and the subsequent accretion gives rise to a luminous transient. A fraction of such tidal disruption events (TDEs) launch collimated relativistic jets (hereafter jetted TDEs; see \citealt{DeColle2020} for a review). 
So far, only four jetted TDE candidates with on-axis jets have been found, including three objects discovered by the hard X-ray burst alert telescope (BAT) on board \swift more than a decade ago, Sw\,J1644+57 \citep{Bloom2011, Burrows2011, Levan2011, Zauderer2011}, Sw\,J2058+05 \citep{Cenko2012, Pasham2015}, Sw\,J1112-82 \citep{Brown2015, Brown2017_J1112}. In contrast to the first three events, AT2022cmc was recently discovered by the Zwicky Transient Facility (ZTF) in the optical band \citep{Andreoni2022, Pasham2023}. 
These objects exhibit rapidly variable, super-Eddington early-time X-ray emission ($>10^{47}\,{\rm erg\,s^{-1}}$) with a power-law secular decline, as well as extremely bright and long-lived radio emission ($>10^{40}\,{\rm erg\,s^{-1}}$). 

Jetted TDEs form a rare class of transients with limited observational data. They are similar to blazars --- active galactic nuclei with powerful jets beamed towards the observer.
However, the broadband spectral energy distribution (SED) of jetted TDEs do not follow the ensemble properties of blazars in the sense that the ratio of X-ray to radio luminosity is extremely high \citep{Cenko2012}. 
In addition, jetted TDEs might be similar to gamma-ray bursts (GRBs, see \citealt{Kumar2015_GRB_review} for a review), as both are triggered by super/hyper-Eddington accretion onto BHs that produce jets. In the standard GRB fireball model, the long-lasting afterglow emission comes from external shocks propagating into the ambient medium, whereas the seconds-long prompt $\gamma$-ray emission comes from energy dissipation (by e.g., internal shocks or magnetic reconnection) in a region closer to the BH \citep{Zhang2018}. 

Among the \swift jetted TDEs, Sw\,J1644+57 is the most well observed event. Evolution of its millimeter (mm) and radio SED over a decade can be well described by synchrotron emission from an outgoing forward shock \citep{Zauderer2011, Berger2012, Zauderer2013, mimica15_swj1644_afterglow, generozov17_jet_afterglows, Eftekhari2018, Cendes2021}, indeed similar to GRB afterglows. As the blast wave travels through the circumnuclear medium (CNM), the shock is decelerated and the CNM density decreases, resulting in the peak of the radio SED moving to lower frequencies over time.

Unlike the better-understood radio emission, the site and radiation mechanism(s) of the bright X-ray emission in jetted TDEs remain actively debated. 
\citet{Burrows2011} showed that the X-ray spectrum of Sw\,J1644+57 is consistent with synchrotron emission of a particle-starved magnetically-dominated jet, whereas \citet{Bloom2011} found acceptable fits with synchrotron self-Compton (SSC) and external inverse Compton (EIC) models.
\citet{Reis2012} detected a 5\,mHz quasi-periodic oscillation in X-ray observations of Sw\,J1644+57, suggesting that the jet production is modulated by accretion variability near the event horizon. \citet{Crumley2016} studied a wide range of emission mechanisms and concluded that the X-ray emission can be produced by either synchrotron emission or EIC scattering off optical photons from the thermal envelope.
\citet{Kara2016} found a blueshifted ($\sim$0.1--0.2$c$) Fe K$\alpha$ line and the associated reverberation lags in the \xmm data. 
Different interpretations for the reflector have been proposed, including a radiation pressure driven sub-relativistic outflow close to the black hole ($\sim 30R_{\rm g}$ where $R_{\rm g}$ is the gravitational radius; \citealt{Kara2016, Thomsen2022}), and a gas layer accelerated by the interaction between the jet X-rays and a thermal envelope ($\sim300R_{\rm g}$; \citealt{Lu2017}). 

AT2022cmc (ZTF22aaajecp) was discovered as a fast optical transient on 2022 February 11 10:42:40 \citep{Andreoni2022_at_report}. Shortly afterwards, it was detected by follow up observations in the radio \citep{Perley2022_GCN31592} and X-ray \citep{Pasham2022_GCN31601} bands. An optical spectrum obtained by ESO's Very Large Telescope reveals host galaxy lines at the redshift of $z=1.193$ \citep{Tanvir2022_GCN31602}. At the cosmological distance, its X-ray and radio luminosities are comparable to Sw\,J1644+47 at similar phases \citep{Yao2022_ATel15230}. Further multi-wavelength follow-up observations revealed the remarkable similarities between AT2022cmc and Sw\,J1644+57, suggesting that AT2022cmc was indeed a jetted TDE \citep{Andreoni2022, Pasham2023, Rhodes2023}. In this paper, we adopt this interpretation.

As the only jetted TDE discovered in the last decade, AT2022cmc offers a great opportunity to address several key questions related to the X-ray emission of jetted TDEs, such as the jet composition, the particle acceleration and energy dissipation processes, and the emission mechanisms. By computing the X-ray power density spectrum, \citet{Pasham2023} demonstrated that the rest-frame systematic X-ray variability timescale is $t_{\rm var}\lesssim 10^3/(1+1.193)$\,s. By requiring that $t_{\rm var}$ exceed the light-crossing time of the Schwarzschild radius of the BH, an upper limit of the BH mass can be derived as $M_{\rm BH}\lesssim 5\times 10^7\,M_\odot$. 
% https://ned.ipac.caltech.edu/level5/Sept18/Rieger/Rieger2.html

\citet{Pasham2023} fitted the radio and soft X-ray SEDs of AT2022cmc with SSC/EIC models, concluding that the relativistic jet exhibits a high ratio of electron-to-magnetic-field energy densities.
In this work, we present \nustar hard X-ray observations, independently analyze the soft X-ray and UV data, and reexamine the broadband SED evolution across nine orders of magnitude in frequency. 
We follow the physical picture outlined in \citet{Andreoni2022} and propose that the observed broken power-law X-ray spectrum can be explained with a synchrotron origin. 

The paper is organized as follows. 
We describe the observations and data reduction in \S\ref{sec:obs}.
In \S\ref{sec:modeling}, we first outline the rationales of treating the broadband radiation with three separate emission components (\S\ref{subsec:SEDfit_outline}), and then perform model fitting on the sub-mm/radio (\S\ref{subsec:radio_syn}), UV/optical (\S\ref{subsec:uvopt_bbody}), and X-ray (\S\ref{subsec:xray_syn}) SEDs.
A discussion is given in \S\ref{sec:discuss}.

Hereafter we use $t_{\rm obs}$ ($t_{\rm rest}$) to denote observer-frame (rest-frame) time relative to the first ZTF detection.  
We adopt a redshift of $z=1.1933$ \citep{Andreoni2022}, a standard $\Lambda$CDM cosmology with $\Omega_{\rm M} = 0.3$, $\Omega_{\Lambda}=0.7$, and $H_0=70\,{\rm km\,s^{-1}\,Mpc^{-1}}$. 
The luminosity distance $d_L=8.22$\,Gpc.
We use UT time and the usual notation $\mathcal{Q}_n = \mathcal{Q}/10^n$, where $\mathcal{Q}$ is in CGS units.
Uncertainties are reported at the 68\% confidence intervals unless otherwise noted, and upper limits are reported at $3\sigma$. 

\section{Observation and Data Analysis} \label{sec:obs}

\begin{deluxetable*}{ccccccccc}[htbp!]
\tablecaption{Log of X-ray Observations Used in Joint Spectral Analysis. \label{tab:nustar}}
\tablehead{
\colhead{Epoch}  
& \colhead{$t_{\rm obs}$} 
& \colhead{$t_{\rm rest}$} 
& \colhead{Mission}  
& \colhead{obsID}  
& \colhead{Exp.} 
& \colhead{Start Time} 
& \colhead{End Time} 
& \colhead{Count Rate [Energy Range]} \\
\colhead{}
& \colhead{(d)}
& \colhead{(d)}
& \colhead{}
& \colhead{}
& \colhead{(ks)}
& \colhead{(UT)}
& \colhead{(UT)}
& \colhead{(\cps [keV])}
}
\startdata
\multirow{3}{*}{1}  & \multirow{3}{*}{7.8} & \multirow{3}{*}{3.6}& \nustar & 80701510002 & 47.9 &  2022-02-18 18:16 & 2022-02-19 18:36  & $0.2167 \pm 0.0022$ [3--27] \\
\cline{4-9} 
 &   & & \multirow{2}{*}{\nicer}    & 4656010102  & 13.5 &  2022-02-18 00:06 & 2022-02-18 22:34 & \multirow{2}{*}{$1.216 \pm 0.010$  [0.3--5]} \\
  &  & & &  4656010103  &  9.6 &  2022-02-18 23:58 & 2022-02-19 23:20   & \\
\hline
\multirow{3}{*}{2}  & \multirow{3}{*}{17.6} & \multirow{3}{*}{8.0}   & \nustar & 90801501002 & 44.5 &  2022-02-28 13:16 & 2022-03-01 10:46  & $0.0899 \pm 0.0015$ [3--24] \\
\cline{4-9}
 & & & \multirow{2}{*}{\nicer} & 4202560109  &  5.9 &  2022-02-28 00:18 & 2022-02-28 23:42   & \multirow{2}{*}{$0.287 \pm 0.009$ [0.3--4]} \\
  & & &  & 5202560101  &  9.2 &  2022-03-01 00:55 & 2022-03-01 22:56   &  \\
\hline
\multirow{2}{*}{3} & \multirow{2}{*}{36.2}  & \multirow{2}{*}{16.5}   & \nustar & 90802306004 & 44.6 &  2022-03-19 03:11 & 2022-03-20 02:56  & $0.0063 \pm 0.0003$ [3--17] \\
\cline{4-9}
  & & & \swift  & 15023014 & 12.5 & 2022-03-18 22:29 & 2022-03-19 22:28 & $0.0111 \pm 0.0010$ [0.3--10]\\
\enddata
\tablecomments{The last column is the mean net count rate within the energy range where the source is above background. For \nustar observations we show the total count rate in the two optical modules (FPMA and FPMB). }
\end{deluxetable*}

All X-ray observations were processed using \texttt{HEASoft} version 6.31.1.
X-ray spectral fitting was performed with \texttt{xspec} (v12.13, \citealt{Arnaud1996}). 
We used the \texttt{vern} cross sections \citep{Verner1996} and the \citet{Anders1989} abundances for the X-ray absorption by the neutral hydrogen column along the line of sight.

\subsection{\nustar} \label{subsec:obs_nustar}

We obtained Nuclear Spectroscopic Telescope ARray (\nustar; \citealt{Harrison2013}) observations under a pre-approved Target of Opportunity (ToO) program (PI: Y. Yao; obsID 80701510002) and Director's Discretionary Time (DDT) programs (PI: Y. Yao; obsIDs 90801501002, 90802306004). The three epochs of observations are summarized in Table~\ref{tab:nustar}.
The first two \nustar observations were conducted jointly with \nicer, and the last \nustar observation was conducted jointly with \swift/XRT.

To generate the first epoch's spectra for the two photon counting detector modules (FPMA and FPMB), source photons were extracted from a circular region with a radius of $r_{\rm src}=45^{\prime\prime}$ centered on the apparent position of the source in both FPMA and FPMB. The background was extracted from a $r_{\rm bkg}=80^{\prime\prime}$ region located on the same detector. 
For the second and third epochs, since the source became fainter, we used $r_{\rm src}=40^{\prime\prime}$ and $r_{\rm src}=35^{\prime\prime}$, respectively. 

All spectra were binned with \texttt{ftgrouppha} using the optimal binning scheme developed by \citet{Kaastra2016}. For the first two \nustar epochs, we further binned the spectra to have at least 20 counts per bin.

\subsection{\nicer} \label{subsec:obs_nicer}
\target was observed by the Neutron Star Interior Composition Explorer (\nicer; \citealt{Gendreau2016}) under ToO (PI: D. R. Pasham) and DDT (PI: Y. Yao) programs from 2022 February 16 to 2022 June 11. 

\begin{figure}[htbp!]
    \centering
    \includegraphics[width=\columnwidth]{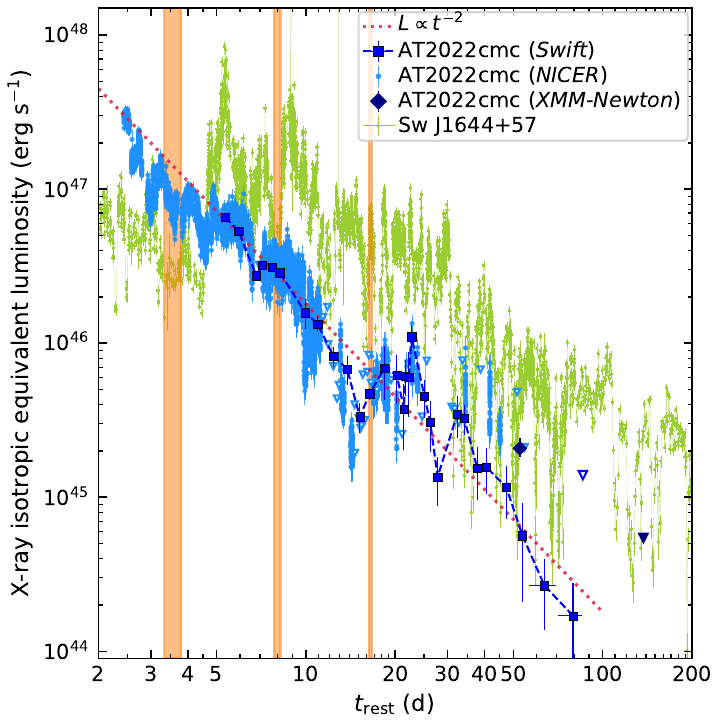}
    \caption{X-ray light curve of \target in the observer-frame 0.3--10\,keV. 
    The vertical bands mark epochs of the three \nustar observations.
    For comparison, we also show the 0.3--10\,keV X-ray light curve of Sw\,J1544+57 \citep{Mangano2016}.}
    \label{fig:xlc}
\end{figure}

First, we ran \texttt{nicerl2} to obtain the cleaned and screened event files, and ran \texttt{nicerl3-lc} to obtain light curves in the  0.3--1\,keV, 1--5\,keV, 0.3--5\,keV and 13--15\,keV bands with a time bin of 30\,s. 
\texttt{nicerl3-lc} estimates the background using a space weather model. 
For good time intervals (GTIs) where more than four focal plane modules (FPMs) were turned off, we scaled the count rate up to an effective area with 52 FPMs. 
We removed four obsIDs where the background count rate is NaN, and removed time bins where the 13--15\,keV count rate is above 0.1\,\cps.
\target was detected above 3$\sigma$ in both 0.3--1\,keV and 1--5\,keV in 39 obsIDs. 
For the remaining 30 obsIDs, we computed the 3$\sigma$ upper limits. 

For each of the 39 obsIDs with significant detections, we ran \texttt{nicerl3-spec} to extract one spectrum using the default parameters. 
Using $C$-statistics \citep{Cash1979}, we fitted each spectrum in the 0.22--15\,keV energy range with a combination of source and background models. 
The source model is an absorbed power-law (i.e., \texttt{tbabs*ztbabs*zashift*(cglumin*powerlaw)} in \texttt{xspec}). 
The background model includes both X-ray and non-X-ray components\footnote{See \url{https://heasarc.gsfc.nasa.gov/docs/nicer/analysis_threads/scorpeon-xspec/} for details.}.
We fixed the Galactic hydrogen-equivalent column density to be $N_{\rm H} = 8.88\times 10^{19}\,{\rm cm^{-2}}$ \citep{HI4PI2016}, and the host absorption to be $N_{\rm H, host}= 10^{21}\,{\rm cm^{-2}}$, which is the best-fit value found in the first joint spectral analysis (see \S\ref{subsubsec:joint1}). 
When $t_{\rm rest} > 10$\,d, the source flux was too faint to provide stringent constraints on both the power-law index and the normalization. Therefore, we further fixed the power-law index $\Gamma = 1.6$. 
Using the best-fit spectral models, we obtained the conversion factors to convert 0.3--5\,keV count rate to 0.3--10\,keV flux (in both observer frame and the rest frame). Figure~\ref{fig:xlc} shows the resulting \nicer light curve.

% \wl{to be continued}

Using observations contemporaneously obtained with the first two \nustar observations (see \S\ref{subsec:obs_nustar}), we also produced two \nicer spectra to be jointly analyzed with the \nustar spectra (see \S\ref{subsubsec:joint1} and \S\ref{subsubsec:joint2}).
The obsIDs of the \nicer data used in this step are shown in Table~\ref{tab:nustar}. 
The source and background spectra were created with \texttt{nibackgen3C50}.
Following the screening criteria suggested by \citet{Remillard2022}, we removed GTIs with \texttt{hbgcut=0.05} and \texttt{s0cut=2.0}, and
added systematic errors of 1.5\% with \texttt{grppha}.

\subsection{\swift/XRT}
\target was observed by the X-Ray Telescope (XRT; \citealt{Burrows2005}) on board \swift following a series of ToO requests (submitted by Y. Yao and D. R. Pasham). All XRT observations were obtained under the photon counting (PC) mode.

We generated the XRT light curve using an automated online tool\footnote{\url{https://www.swift.ac.uk/user_objects}} \citep{Evans2007, Evans2009}. 
For data at $t_{\rm rest}<19$\,d, we binned the light curve by obsID.
For data at $t_{\rm rest}>19$\,d, we used dynamic binning to ensure a minimum of five counts per bin. 
Using the same tool, we also created three stacked XRT spectra for data at $t_{\rm rest}<9$\,d, $9<t_{\rm rest}<19$\,d, and $t_{\rm rest}>19$\,d. We then fitted the three spectra using the same absorbed power-law model as described in \S\ref{subsec:obs_nicer}.
From the best-fit models, we obtained conversion factors to convert 0.3--10\,keV net count rate to 0.3--10\,keV flux (in both observer frame and the rest frame). 
The XRT light curve is shown in Figure~\ref{fig:xlc}.

To generate an XRT spectrum for obsID 15023014 (to be jointly analyzed with the third \nustar epoch), we processed the data using \texttt{xrtproducts}. We extracted source photons from a circular region with a radius of $r_{\rm src} = 30^{\prime\prime}$, and background photons from eight background regions with $r_{\rm bkg} = 25^{\prime\prime}$ evenly spaced at $80^{\prime\prime}$ from \target. 
The spectrum was first binned with \texttt{ftgrouppha} using the optimal binning scheme \citep{Kaastra2016}, and then further binned to have at least one count per bin.

\subsection{\xmm}
AT2022cmc was observed twice by \xmm as part of our GO program (PI: S.~Gezari, obsIDs 0882591301, 0882592101).
The first observation took place on 2022 June 6 and lasted $\sim\! 18$\,ks, while the second observation occurred on 2022 December 9 and lasted $\sim\! 21$\,ks. 
Since the pn instrument of the EPIC camera has a larger effective area than MOS1 and MOS2, we only analyzed the pn data. 
The raw data files were processed using the \texttt{epproc} task. 
Following the \xmm data analysis guide, to check for background activity and generate GTIs, we manually inspected the background light curves in the 10--12\,keV band. 
The source was detected in the first epoch, but not in the second one. 

For the first epoch, source photons were extracted from a circular region with $r_{\rm src}=33^{\prime\prime}$ centered on the position of the source. 
The background was extracted from a $r_{\rm bkg} = 45^{\prime\prime}$ region on the same detector. 
The observation data files were reduced using the \xmm\ Standard Analysis Software \citep[SAS;][]{Gabriel_04}. The ARFs and RMF files were created using the \texttt{arfgen} and \texttt{rmfgen} tasks, respectively. 
The resulting EPIC-pn spectrum has $\sim\!200$ background subtracted counts, at a rate of $\sim \!0.023$\,count\,s$^{-1}$. The spectrum was binned using the optimal binning criteria \citep{Kaastra2016}, ensuring that each bin had at least one count.

For the second epoch, we ran the \texttt{eregionanalyse} task using the same apertures as in the first epoch, and obtained a background-subtracted 3$\sigma$ upper limit of $\sim\!0.006$\,count\,s$^{-1}$.

\begin{figure}[htbp!]
    \centering
    \includegraphics[width=\columnwidth]{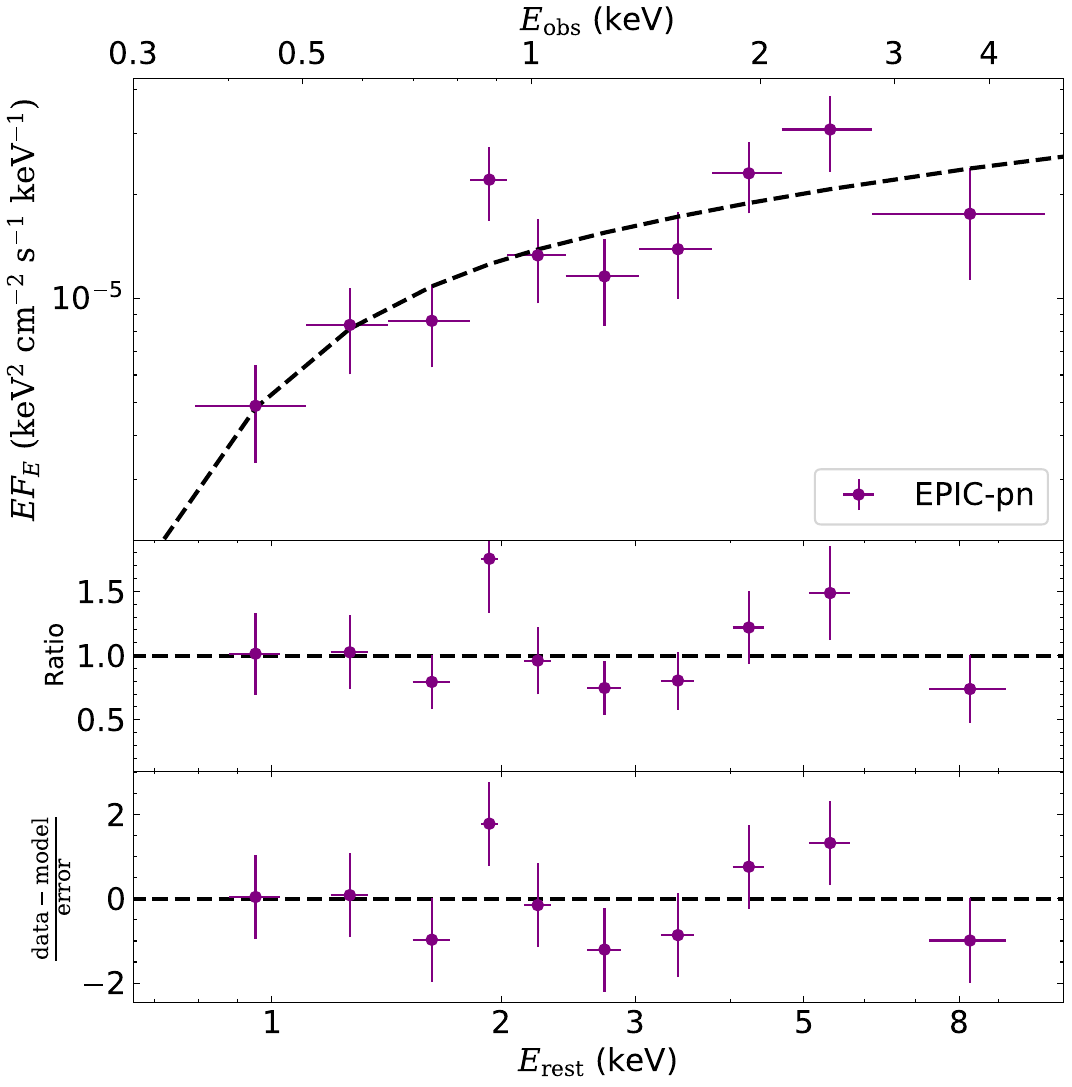}
 \caption{The \xmm EPIC-pn X-ray spectrum at $t_{\rm rest}\approx 52.6$\,d.}
    \label{fig:xmm_spec}
\end{figure}

For the first epoch \xmm EPIC-pn data, we selected the energy range of 0.35--4.5\,keV, where the source spectrum dominates over the background. The data was modeled with an absorbed power-law using $C$-statistics.
The best-fit model parameters are $N_{\rm H, host} = 0.22_{-0.14}^{+0.31}\times 10^{22}\,{\rm cm^{-2}}$, $\Gamma = 1.65\pm 0.14$, and $cstat/dof = 11/8$. 
Figure~\ref{fig:xmm_spec} displays the spectrum along with the best-fitting model.
The X-ray luminosity at this first epoch is $L_{\rm X} = 2.08^{+0.35}_{-0.25}\times 10^{45}\,{\rm erg\,s^{-1}}$. 
Assuming the same spectrum we estimate the 3$\sigma$ upper-limit luminosity of the second epoch to be $L_{\rm X} < 5.45\times 10^{44}\,{\rm erg\,s^{-1}}$.

\subsection{Joint X-ray Spectral Analysis} \label{subsec:xray_spec}
\begin{figure}[htbp!]
    \centering
    \includegraphics[width=\columnwidth]{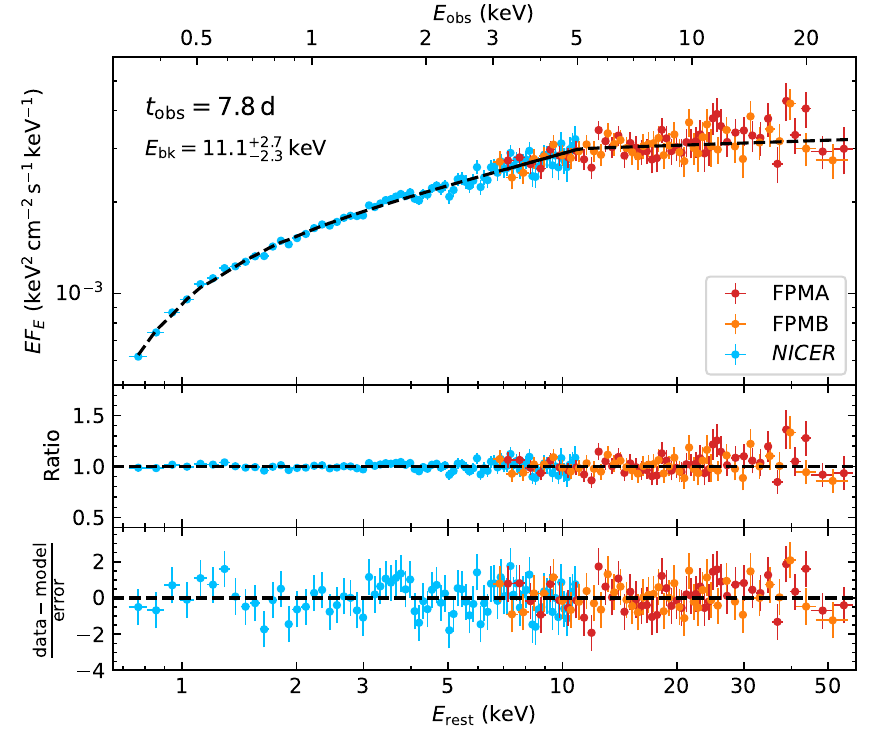}
    \includegraphics[width=\columnwidth]{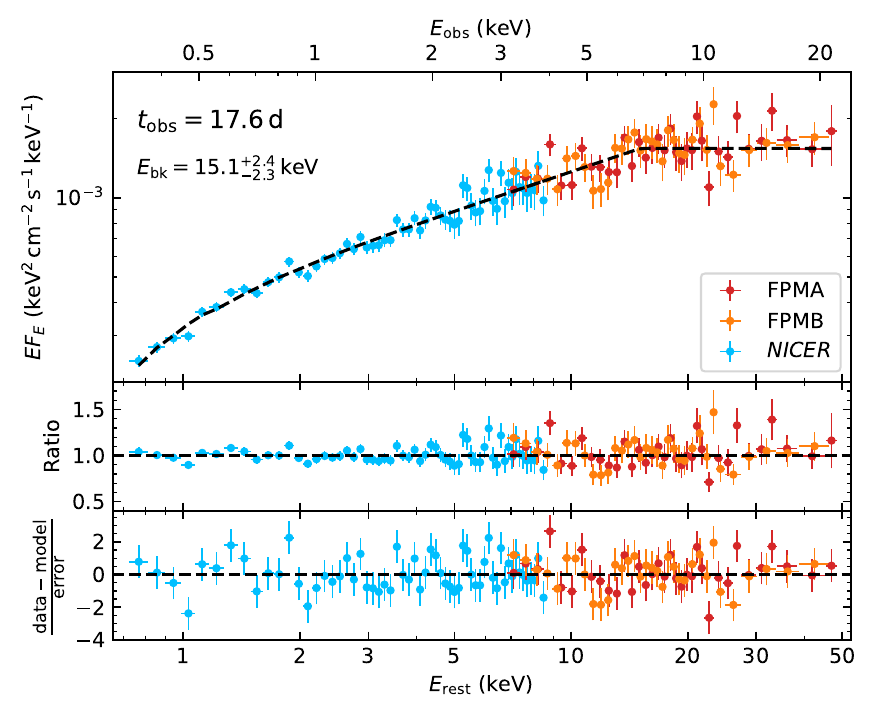}
    \includegraphics[width=\columnwidth]{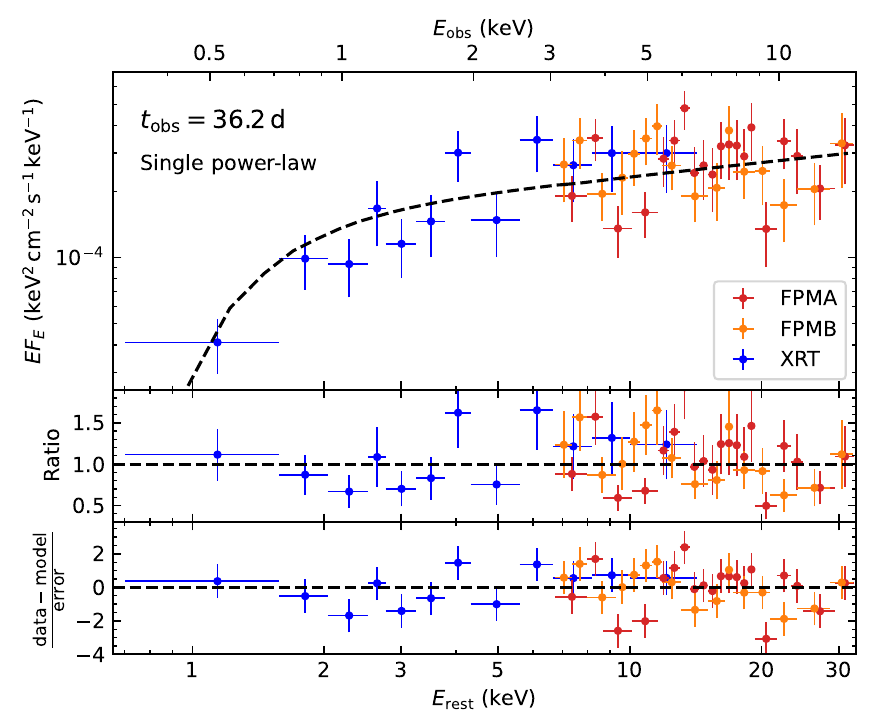}
    \caption{From top to bottom: the X-ray spectrum at three epochs. 
    \nustar/FPMB, \nicer, and XRT data have been divided by $C_{\rm FPMB}$, $C_{\rm NICER}$, and $C_{\rm XRT}$ respectively.
    The data have been rebinned for visual clarity.}
    \label{fig:joint_spec}
\end{figure}

Here performed joint spectral analysis between \nustar and soft X-ray observations (\nicer or XRT). 
Data were fitted with $\chi^2$-statistics for the first two epochs, and with $C$-statistics for the third epoch. 
Uncertainties of X-ray model parameters are reported at the 90\% confidence level. 
For all models described below, we included a calibration coefficient (\texttt{constant}; \citealt{Madsen2017}) between FPMA, FPMB, and \nicer (or XRT), with $\mathcal{C}_{\rm FPMA}$ fixed at one.

\subsubsection{Epoch 1} \label{subsubsec:joint1}
We chose energy ranges where the source spectrum dominates over the background. 
For \nicer we used 0.3--5.0\,keV; For \nustar FPMA and FPMB we used 3--27\,keV. 
Fitting the data with a single power-law results in a poor fit with a $\chi^2$ over degrees of freedom (\textit{dof}) of $302/206 = 1.47$. Replacing the single power-law with a broken power-law (\texttt{bknpower}) gives a good fit. This model assumes that the photon energy distribution takes the form $n(E)dE \propto E^{-\Gamma_1}$ below a break energy $E_{\rm bk}$, and that $n(E)dE \propto E^{-\Gamma_2}$ where $E > E_{\rm bk}$.

The best-fit model is presented in the top panel of Figure~\ref{fig:joint_spec}. 
The best-fit parameters are: 
$C_{\rm FPMB} = 0.99\pm0.03$, 
$C_{\nicer} = 1.11\pm0.05$, 
$N_{\rm H, host} = 1.03_{-0.10}^{+0.11}\times 10^{21}\,{\rm cm^{-2}}$, 
$\Gamma_1 = 1.66\pm0.02$, 
$\Gamma_2 = 1.96_{-0.05}^{+0.08}$, 
$E_{\rm bk} = 11.1_{-2.3}^{+2.7}$\,keV, 
and $\chi^2/dof = 174/204$.
The isotropic-equivalent 0.5--50\,keV X-ray luminosity is $L_{\rm X} = (1.30\pm0.03)\times 10^{47}\,{\rm erg\,s^{-1}}$. 

\subsubsection{Epoch 2} \label{subsubsec:joint2}
We chose energy ranges where the source spectrum dominates over the background. 
For \nicer we used 0.3--4.0\,keV; For \nustar FPMA and FPMB we used 3--24\,keV.
Similar to what was found in \S\ref{subsubsec:joint1}, a single power-law leaves an unacceptable $\chi^2 / dof$ of $204/150=1.36$, whereas a broken power-law describes the data much better (see the middle panel of Figure~\ref{fig:joint_spec}).

The best-fit model parameters are: 
$C_{\rm FPMB} = 1.00_{-0.05}^{+0.06}$, 
$C_{\nicer} = 0.86\pm0.06$, 
$N_{\rm H, host} = 0.55_{-0.22}^{+0.23}\times 10^{21}\,{\rm cm^{-2}}$, 
$\Gamma_1 = 1.51\pm0.04$, 
$\Gamma_2 = 2.00_{-0.12}^{+0.15}$, 
$E_{\rm bk} = 15.1_{-2.3}^{+2.4}$\,keV, 
and $\chi^2/dof = 146/148$.
The isotropic-equivalent 0.5--50\,keV X-ray luminosity is $L_{\rm X} = (0.60\pm0.02)\times 10^{47}\,{\rm erg\,s^{-1}}$. 

\subsubsection{Epoch 3} \label{subsubsec:joint3}
We chose energy ranges where the source spectrum dominates over the background. 
For XRT we used 0.3--10.0\,keV. For \nustar FPMA and FPMB we used 3--15\,keV. 
Compared with the previous two \nustar epochs, \target has become much fainter at this observing epoch. Both a single power-law and a double power-law give acceptable fits. 

First, we model the X-ray spectrum with a double power-law with $\Gamma_1 = 1.5$ and $\Gamma_2 = 2.0$ (similar to the previous two epochs). 
The best-fit model parameters are: 
$C_{\rm FPMB} = 0.92_{-0.13}^{+0.15}$, 
$C_{\rm XRT} = 0.81_{-0.25}^{+0.19}$,
$N_{\rm H, host} = 0.26_{-0.25}^{+0.37}\times 10^{21}\,{\rm cm^{-2}}$, 
$E_{\rm bk} = 11.7_{-6.5}^{+3.6}$\,keV, 
and $cstat/dof = 140/98$.
Next, we model the X-ray spectrum with a single power-law.
The best-fit model parameters are: 
$C_{\rm FPMB} = 0.92_{-0.13}^{+0.15}$, 
$C_{\rm XRT} = 0.68_{-0.17}^{+0.22}$,
$N_{\rm H, host} = 0.56_{-0.36}^{+0.54}\times 10^{21}\,{\rm cm^{-2}}$, 
$\Gamma = 1.79_{-0.17}^{+0.18}$\,keV, 
and $cstat/dof = 129/98$.
The bottom panel of Figure~\ref{fig:joint_spec} shows the single power-law fit, which is favored by the fit statistics. 
The X-ray luminosity at this epoch is $L_{\rm X} = (0.12\pm0.01)\times 10^{47}\,{\rm erg\,s^{-1}}$. 

%\subsection{$\gamma$-ray}
%Since no $\gamma$-ray detection has been reported from the Large Area Telescope (LAT; \citealt{Atwood2009}) on board the \textit{Fermi} satellite, we assume a nominal LAT upper limit at 0.1--10\,GeV to be $<2.4\times 10^{-11}\,{\rm erg\,s^{-1}\,cm^{-2}}$ \citep{Peng2016}. 

\subsection{UV and optical} \label{subsec:uvopt_obs}

\begin{figure}[htbp!]
    \centering
    \includegraphics[width=\columnwidth]{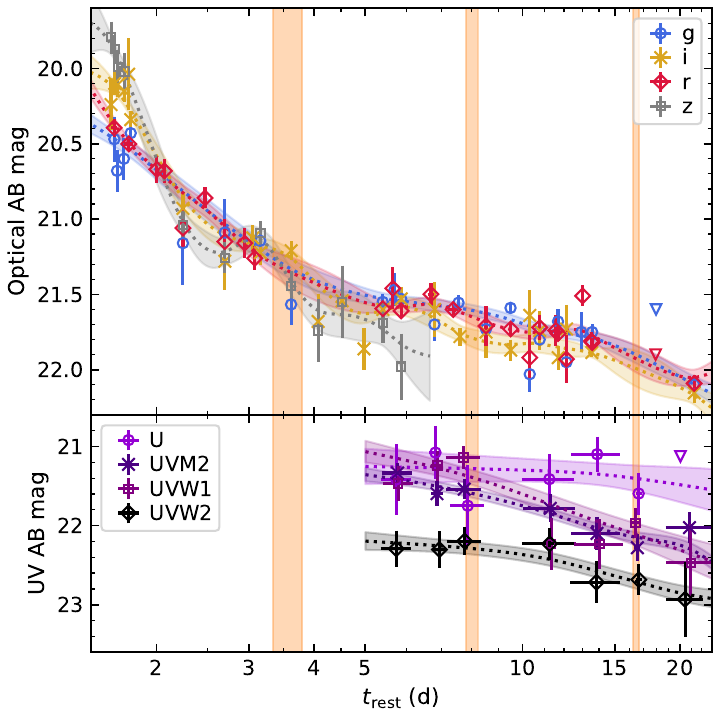}
    \caption{
    Optical and UV photometry of \target. 
    The transparent lines are simple Gaussian process fits in each filter, where the width of the lines represent 1$\sigma$ model uncertainties.
    The vertical bands mark periods of the three \nustar observations. }
    \label{fig:uvolc}
\end{figure}

The top panel of Figure~\ref{fig:uvolc} shows the optical data reported in \citet[][Supplementary table 1]{Andreoni2022}.
For UV data taken by the Ultra-Violet/Optical Telescope (UVOT; \citealt{Roming2005}) on board \swift, we stacked a few adjacent obsIDs with \texttt{uvotimsum} to improve the sensitivity, and performed photometry on the stacked images with \texttt{uvotsource}. The bottom panel of Figure~\ref{fig:uvolc} shows the results. 

We note that the UV and optical photometry exhibits short-timescale ($\sim \! {\rm hr}$--day) wiggles (either due to intrinsic stochastic variability or underestimated systematic uncertainties across multiple instruments). Therefore, to capture the general trend of the photometric evolution, we fit the UV and optical data in each filter with Gaussian process models, following the same procedures described in \citet{Yao2022}. We then infer the photometry at the \nustar observing epochs using the best-fit models (shown as the transparent lines in Figure~\ref{fig:uvolc}).

\subsection{Radio/sub-mm} \label{subsec:radio_obs}

\begin{figure}[htbp!]
    \centering
    \includegraphics[width=\columnwidth]{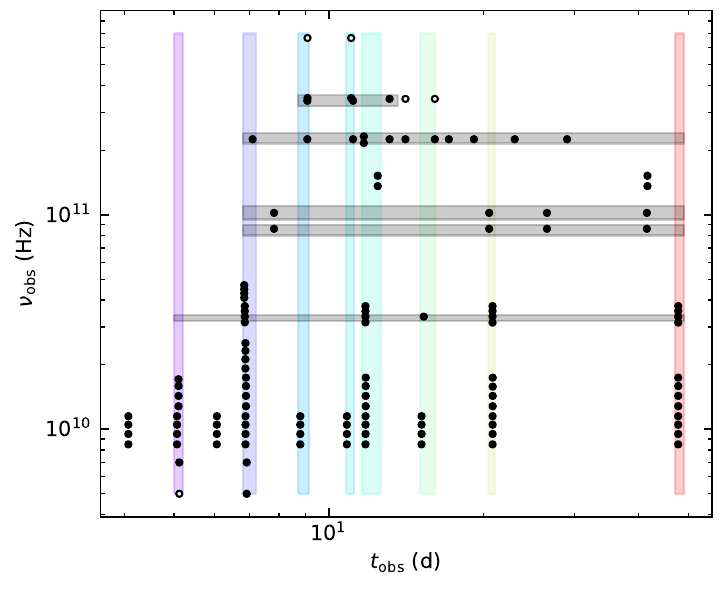}
    \caption{Radio/mm observing time and frequency reported in \citet{Andreoni2022}.
    Detections are shown in solid circles and upper limits are shown as hollow circles.
    We fitted the light curves in five frequencies (33.5, 86, 102, 225, and 347\,GHz) over the time range indicated by the vertical grey bands to infer the flux densities at the epochs with dense low-frequency coverage (marked by the vertical colored bands). 
    \label{fig:time_freq}}
\end{figure}

\begin{figure}[htbp!]
    \centering
    \includegraphics[width = \columnwidth]{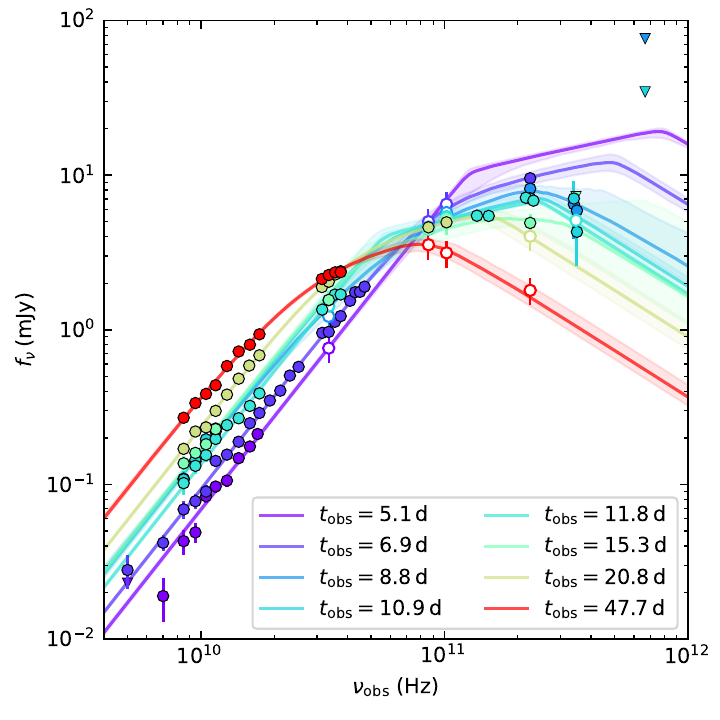}
    \caption{Radio SED of \target, overplotted with the best-fit afterglow models (see \S\ref{subsec:radio_syn}). 
    Solid circles (with the black edge color) are measurements;
    Hollow circles are inferred from Gaussian process fitting of radio light curves in the corresponding frequencies. 
    \label{fig:radio_sed}}
\end{figure}

In this work, we analyze radio and sub-mm observations of \target reported in \citet{Andreoni2022}. Figure~\ref{fig:xlc} shows the time and frequency of the observations. We prepare eight epochs (indicated by the vertical colored bands) of SEDs with good frequency sampling to be analyzed in \S\ref{sec:modeling}. Since the sub-mm observations were much sparser, we fitted the light curves at five frequencies (indicated by the horizontal grey bands) with Gaussian process models to infer the high-frequency flux densities. The resulting SED at the eight epochs are shown in Figure~\ref{fig:radio_sed}. 

\section{Broadband SED Modeling} \label{sec:modeling}

\begin{figure*}[htbp!]
    \centering
    \includegraphics[width=\textwidth]{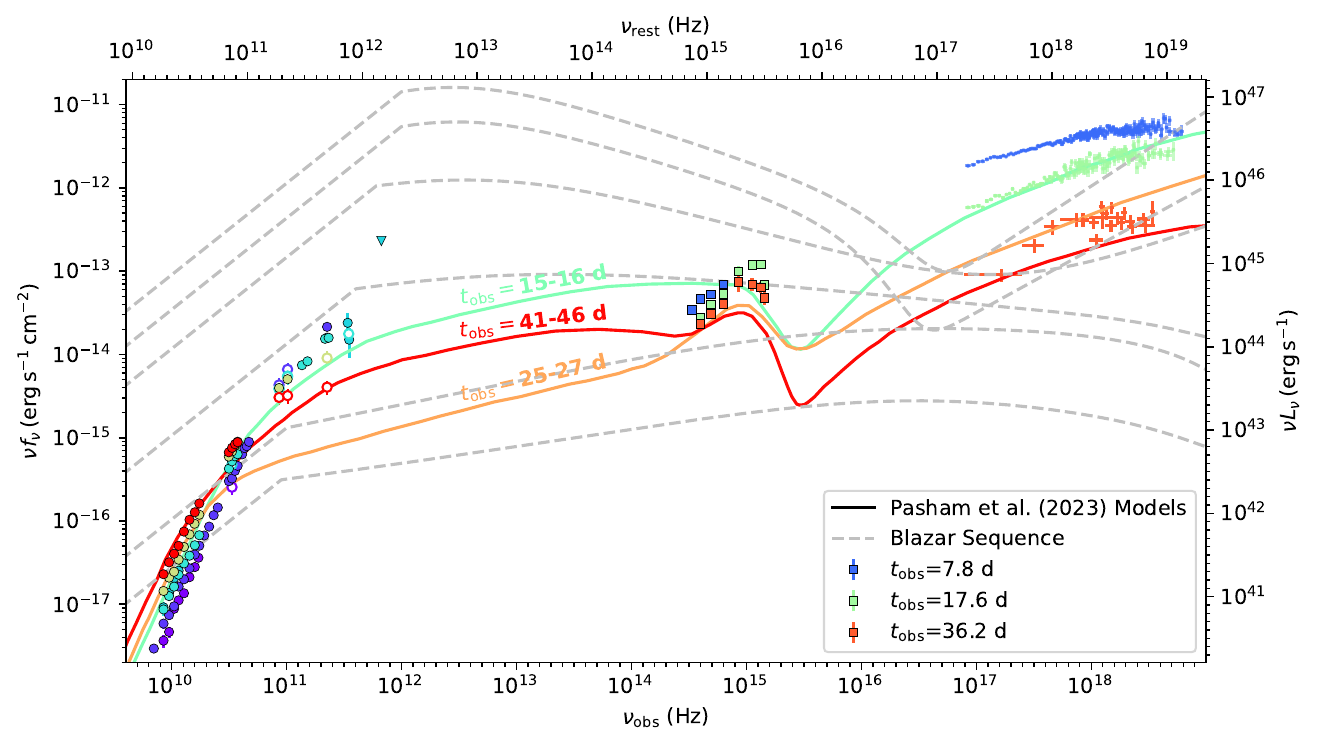}
    \caption{Broadband SEDs of AT2022cmc, as compared to the best-fit synchrotron+blackbody+SSC models (solid lines) from \citet[][their Fig.~3]{Pasham2023}.
    We show the Galactic and host galaxy absorption corrected X-ray spectra (see \S\ref{subsec:xray_spec}), the Galactic extinction corrected UV/optical data at the \nustar observing epochs (solid squares, see \S\ref{subsec:uvopt_obs}), and the observed radio/sub-mm data at eight epochs from $t_\textrm{obs}=5.1$\,d to 47.7\,d (circles, see \S\ref{subsec:radio_obs} and Figure~\ref{fig:radio_sed}). The dashed grey lines show the blazar sequence from \citet[][their Tab.~1]{Ghisellini2017}. \\
    \label{fig:22cmc_sed_p23}}
\end{figure*}

\subsection{Preliminary Considerations} \label{subsec:SEDfit_outline}
%%% properties
Since \citet{Pasham2023} is the only previous work that have modeled the radio-to-X-ray SED of \target, we briefly summarize their results here.

\citet{Pasham2023} consider the scenario where the X-ray and radio photons are emitted from the same region at the jet front, whereas the UV/optical emission originates from a quasi-spherical thermal envelope (modeled with a blackbody). 
In their model 1 (synchrotron+SSC), the jet radio synchrotron photons are inverse Compton scattered by relativistic electrons to produce the X-ray emission; in their model 2 (synchrotron+EIC), seed photons from a thermal envelope outside the jet are inverse Compton scattered to produce the X-rays. 
SED fitting was performed at three epochs with good multi-wavelength coverage ($t_{\rm obs}=15$--16\,d, 25--27\,d, and 41--46\,d) using the \texttt{BHjet} code developed by \citet{Lucchini2022}. 
\citet{Pasham2023} find that model 1 was favored over model 2. 

Figure~\ref{fig:22cmc_sed_p23} displays the best-fit synchrotron+blackbody +SSC models from \citet{Pasham2023}. Although the 15--16\,d model is fitted to data obtained close in time to the second \nustar epoch ($t_{\rm obs}=17.6$\,d), it fails to match the observed optical spectral slope or reproduce the broken power-law shape in the X-ray band. 
Moreover, both the 25--27\,d model and the 41--46\,d model significantly under-predict the 30--300\,GHz flux, which likely results from the fact that sub-mm data was not included in the SED fitting. 
Notably, the models are in conflict with the observed 100\,GHz light curve of AT2022cmc, which exhibits a slight monotonic decline from $t_{\rm obs}=16$\,d to 60\,d (see \citealt[][Fig.~1]{Andreoni2022}). 

A novel result from our joint \nicer and \nustar observations is that the X-ray spectrum exhibits a relatively sharp break (at least in the first two epochs, see Figure~\ref{fig:joint_spec}), whereas the spectra produced by the SSC process are quite smooth \citep{Ghisellini2013, Zhang2018}.

Given the aforementioned issues of the synchrotron+SSC models, hereafter we consider an alternative scenario where the X-ray and radio photons arise from two separate regions, akin to the prompt and afterglow emitting sites observed in GRBs. 
This physical picture is motivated by the following two arguments. First, the external shock model developed for GRB afterglows has been successfully applied to the radio/mm observations of Sw\,J1644+57 \citep{Zauderer2011, Berger2012, Zauderer2013, mimica15_swj1644_afterglow, Eftekhari2018, Cendes2021}. 
Second, the observed host-frame X-ray variability timescale places an upper limit on the size of the X-ray emitting region, $R_{\rm X}\lesssim c \Gamma_{\rm j}^2 t_{\rm var}=1.2\times 10^{16}(\Gamma_{\rm j}/30)^2$\,cm; and the external shock modeling of the radio SEDs places a lower limit on the size of the radio emitting region, $R_{\rm radio}\gtrsim 10^{17}$\,cm (see details in \S\ref{subsec:radio_syn}). The fact that $R_{\rm X}$ and $R_{\rm radio}$ are not consistent indicates that X-ray and radio photons are coming from different regions. 

A similar interpretation has also been adopted by several previous works to explain Sw\,J1644+57 \citep{Crumley2016, Lu2017} and AT2022cmc \citep{Andreoni2022, Matsumoto2023}. 
For the X-ray emission of AT2022cmc, we explore the possibility of a pure synchrotron origin, which is the leading emission mechanism for the GRB prompt emission \citep{Oganesyan2019, Zhang2020}.

\subsection{Sub-mm/Radio: External Shock} \label{subsec:radio_syn}

We use the standard external shock afterglow model \citep{Sari1998, Granot2002} to fit the sub-mm and radio SEDs.
Here $\Gamma = 1 / \sqrt{1 - \beta^2}$ is the bulk Lorentz factor of the emitting region. 
The electrons in the shock are accelerated into a power-law distribution, 
$N(\gamma_{\rm e}) \propto \gamma_{\rm e}^{-p}$ for $\gamma_{\rm e} \geq \gamma_{\rm m}$, 
where $\gamma_{\rm m}$ is the minimum Lorentz factor of the relativistic electrons.
A fraction $\epsilon_{\rm e}$ of the shock energy goes into electrons and a fraction $\epsilon_{\rm B}$ of shock energy goes into magnetic energy density.
The critical electron Lorentz factor at which synchrotron cooling time equals to the dynamical time is $\gamma_{\rm c}$.
The characteristic synchrotron frequencies for electrons with $\gamma_{\rm e} =\gamma_{\rm m}$ and $\gamma_{\rm e} =\gamma_{\rm c}$ are denoted as $\nu_{\rm m}$ and $\nu_{\rm c}$, respectively.
The self-absorption frequency is denoted as $\nu_{\rm a}$, below which the system is optically thick to its own synchrotron emission. 

From the observed radio spectra, we infer that the system is in the slow cooling regime, i.e., $\nu_{\rm m} < \nu_{\rm c}$. 
In this regime, one can use spectrum 1 and spectrum 2 of \citet{Granot2002}. 
At sufficiently early times, $\nu_{\rm a} \ll \nu_{\rm m}$, the synchrotron spectrum is given by 
\begin{align}
    F_{\nu, 1}(\nu)  = & F_\nu (\nu_{\rm a}) \left[ \left( \frac{\nu}{\nu_{\rm a}}\right)^{-s_1 \beta_1} +  \left( \frac{\nu}{\nu_{\rm a}}\right)^{-s_1 \beta_2} \right]^{-1/s_1}  \notag \\
    & \times \left[ 1 +  \left( \frac{\nu}{\nu_{\rm m}}\right)^{s_2 (\beta_2 - \beta_3)} \right]^{-1/s_2} 
\end{align}
where $s_1$ and $s_2$ are smoothing parameters, $\beta_1 = 2$, $\beta_2 = 1/3$, $\beta_3 = (1-p)/2$ are the power-law indices of each segment. At late times, $\nu_{\rm m} \ll \nu_{\rm a}$, and we have
\begin{align}
    F_{\nu, 2} (\nu)  = & F_\nu (\nu_{\rm m}) \left[ \left( \frac{\nu}{\nu_{\rm m}}\right)^2 e^{-s_4 \left(\nu/\nu_{\rm m} \right)^{2/3} } + \left( \frac{\nu}{\nu_{\rm m}}\right)^{5/2} \right] \notag \\
    & \times \left[ 1 + \left(\frac{\nu}{\nu_{\rm a}} \right)^{s_5(\beta_2 - \beta_3) } \right]^{-1/s_5}
\end{align}
where $s_4$ and $s_5$ are smoothing parameters, $\beta_2 = 5/2$, and $\beta_3 = (1-p)/2$. 
To smoothly connect the evolution in the two phases, we follow \citet{Berger2012} and use a weighted average
\begin{align}
    F_\nu (\nu) = \frac{w_1 F_{\nu, 1} + w_2 F_{\nu, 2}}{w_1 + w_2}
\end{align}
where $w_1 = (\nu_{\rm m} / \nu_{\rm a})^2$ and $w_2 = (\nu_{\rm a} / \nu_{\rm m})^2$. 

As the optically thin part of the spectrum is not well sampled at early times, hereafter we assume $p=3$, which is the typical value found in Sw\,J1644+57 \citep{Cendes2021}. 
We perform the fit using the Markov chain Monte Carlo (MCMC) approach with \texttt{emcee} \citep{Foreman-Mackey2013}. 
The best-fit models are shown in Figure~\ref{fig:radio_sed}.
For all radio epochs analyzed in this work, $w_1 \gg 1$, $F_{\nu}(\nu)\approx F_{\nu, 1}$, the observed peak frequency $\nu_{\rm p} = \nu_{\rm m}$, and the observed peak specific flux $F_{\rm p} = F_\nu(\nu_{\rm p})$.

Using equipartition analysis in the relativistic regime \citep{BarniolDuran2013_equipartition}, we computed the equipartition radius
\begin{align}
    R_{\rm eq} \approx (1.7\times 10^{17}\,{\rm cm}) \left[ \frac{F_{p, {\rm mJy}}^{8/17} d_{L, 28}^{16/17} \eta^{35/51}  }{ \nu_{p, 10} (1+z)^{25/17} } \right] \frac{\Gamma^{10/17}}{ f_A^{7/17} f_{V}^{1/17} },
\end{align}
and the minimal total energy
\begin{align}
    E_{\rm eq} \approx (2.5\times 10^{49}\,{\rm erg}) \left[ \frac{F_{p, {\rm mJy}}^{20/17} d_{L, 28}^{40/17} \eta^{15/17} }{\nu_{p, 10} (1+z)^{37/17} } \right] \frac{ f_{V}^{6/17} }{ f_A^{9/17} \Gamma^{26/17}}.
\end{align}
Here $f_V$ and $f_A$ are geometry factors, and $\eta = \nu_{\rm m} / \nu_{\rm a}$. 
We consider a narrow jet with a half-opening angle of $\theta_{\rm j}= 0.1 < 1/\Gamma$, such that $f_A = f_V = (\theta_{\rm j} \Gamma)^2$. 

Following \citet{BarniolDuran2013_J1644, BarniolDuran2013_equipartition} and \citet{Eftekhari2018}, we assume $\epsilon_{\rm e} = 0.1$, $\epsilon_{\rm B} = 10^{-3}$, and that the kinetic energy of hot protons is 10 times more than the electrons. 
Defining $\xi \equiv 1+\epsilon_{\rm e} = 11$, the equipartition radius will be increased by a factor of $\xi^{1/17} =1.15$ and the total minimal energy will be increased by a factor of $\xi^{11/17}=4.72$. 
Defining $\epsilon \equiv (\epsilon_{\rm B}/\epsilon_{\rm e})/(6/11)$, the actual radius $R$ corresponding to the minimum energy is different from $R_{\rm eq}$ by a multiplicative factor of $\epsilon^{1/17}=0.79$ and the total energy $E_{\rm T}$ is greater than $E_{\rm eq}$ by a multiplicative factor of $(11/17)\epsilon^{-6/17}+(6/17)\epsilon^{11/17}=2.68$.

The magnetic field in the source frame is
\begin{align}
    B = (1.3\times 10^{-2}\,{\rm G}) \left[ \frac{ \nu_{p, 10}^5 (1+z)^7 }{ F_{p, {\rm mJy}}^{2} d_{L, 28}^{4} \eta^{10/3} } \right] \frac{f_A^2 R_{17}^4}{ \Gamma^3 }.
\end{align}
$\Gamma$ is related to the emitting radius $R$ by
\begin{align}
    t \approx \frac{R(1-\beta) (1+z)}{\beta c}.
\end{align}

\begin{figure}[htbp!]
    \centering
    \includegraphics[width=\columnwidth]{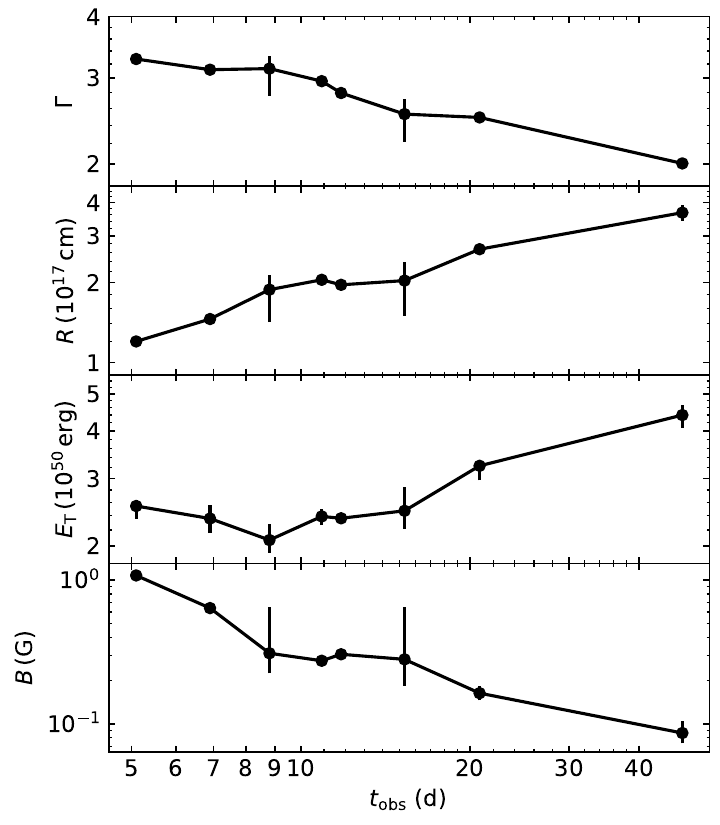}
    \caption{Evolution of physical properties inferred by fitting the sub-mm/radio SEDs. \label{fig:FS_pars}}
\end{figure}

\begin{figure*}[htbp!]
    \centering
    \includegraphics[width=\textwidth]{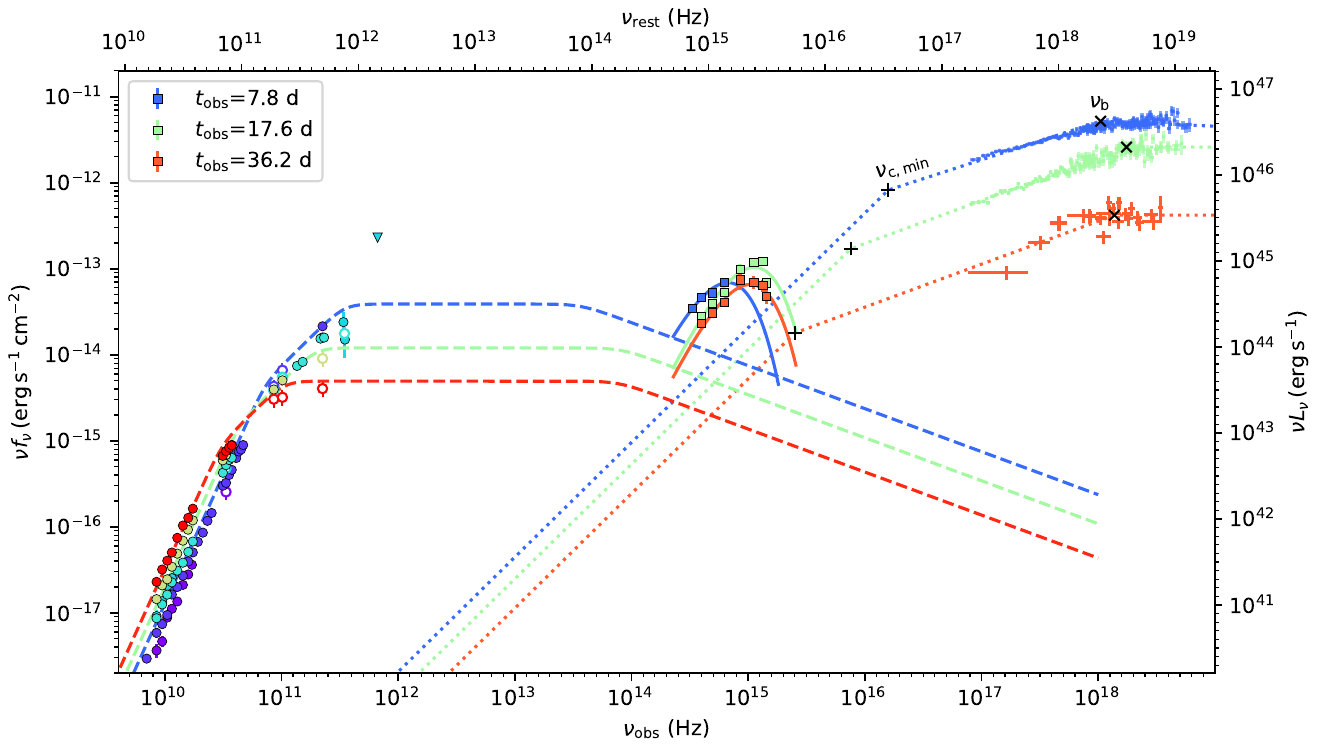}
    \caption{Broadband SEDs of AT2022cmc, as compared to our 3-component best-fit models at the \nustar observing epochs. The models include the sub-mm/radio synchrotron component (dashed lines, see \S\ref{subsec:radio_syn}), the UV/optical blackbody component (solid lines, see \S\ref{subsec:uvopt_bbody}), and an example of the X-ray synchrotron component (dotted lines, \S\ref{subsec:xray_syn}).
    The minimum cooling frequency $\nu_{\rm c, min}$ and the break frequency $\nu_{\rm b}$ are marked by the plus sign and cross sign, respectively. Data are the same as shown in Figure~\ref{fig:22cmc_sed_p23}. 
    \label{fig:22cmc_sed}}
\end{figure*}

Figure~\ref{fig:FS_pars} shows the evolution of $\Gamma$, $R$, $E_{\rm T}$, and $B$. 
Similar values have also been obtained with afterglow model fitting performed by \citet{Matsumoto2023}. 
The cooling frequency $\nu_{\rm c}$ lie in the infrared band.
The best-fit afterglow models at the three \nustar observing epochs are shown as the dashed lines in Figure~\ref{fig:22cmc_sed}.

\subsection{UV/optical: Thermal Envelope} \label{subsec:uvopt_bbody}

\begin{figure}[htbp!]
    \centering
    \includegraphics[width=\columnwidth]{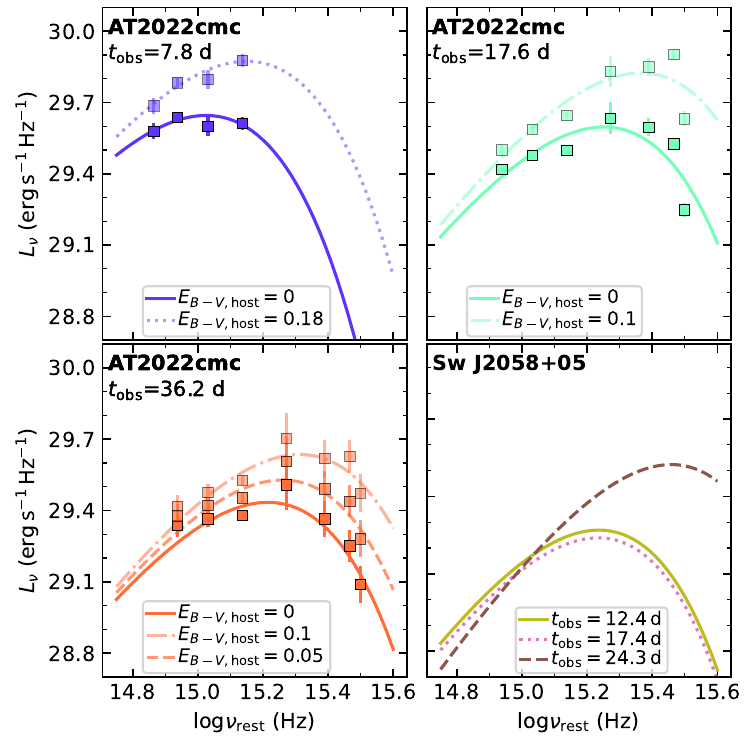}
    \caption{UV/optical SEDs of AT2022cmc at three epochs, overplotted with the best-fit blackbody models assuming various values of host extinction $E_{B-V,{\rm host}}$. The lower right panel shows best-fit blackbody models of Sw\,J2058+05 \citep[][their Tab.~5]{Pasham2015}. \label{fig:blackbody_SED}}
\end{figure}

We analyze the Galactic extinction corrected UV and optical photometry, take $E_{B-V, \rm MW}=0.0095$ \citep{Schlafly2011}, assume $R_V=3.1$, and adopt the reddening law from \citep{Cardelli1989}.  

First, we assume negligible host extinction. The optical ($griz$) spectral slopes $\alpha$ (for $f_\nu \propto \nu^{\alpha}$) at $t_{\rm obs}=7.8$, 17.6, and 36.2\,d are $0.06\pm0.15$, $0.40\pm0.143$, and $0.22\pm0.26$, respectively. 
The optical emission is therefore not an extension of the radio/mm synchrotron SED.
% consistent with the analysis in \S\ref{subsec:radio_syn}.
In the UV bands, the spectral slope at the second and third \nustar epochs are $-1.35\pm0.31$ and $-1.68\pm0.54$. 
Following \citet{Andreoni2022} and \citet{Pasham2023}, we fit the UV/optical SED of AT2022cmc with a blackbody function. 
The best-fit blackbody radius, temperature, and luminosity are presented in Table~\ref{tab:model_pars}.
The derived parameters can be compared with Sw\,J2058+05, which is the only previously known on-axis jetted TDE with early-time multi-band optical photometry and negligible/small host extinction. 
%We note that Sw\,J1644+57 was much too absorbed to get anything meaningful out of the UV/optical
AT2022cmc has blackbody temperatures that are similar to Sw\,J2058+05 and blackbody radii that are a factor of $\sim3$ greater than that obsevred in Sw\,J2058+05 (see Figure~\ref{fig:blackbody_SED}). 

\begin{deluxetable}{ccccc}[htbp!]
\tablecaption{Best-fit Parameters of the Thermal Envelope. \label{tab:model_pars}}
\tablehead{
\colhead{$t_{\rm obs}$} 
& \colhead{$E_{B-V,{\rm host}}$} 
& \colhead{${\rm log}T_{\rm bb}$} 
& \colhead{${\rm log}R_{\rm bb}$} 
& \colhead{${\rm log}L_{\rm bb}$} \\
\colhead{(d)}
& \colhead{(mag)}
& \colhead{(K)}
& \colhead{(cm)}
& \colhead{(erg\,s$^{-1}$)}
}
\startdata
\multirow{2}{*}{7.8} & 0 & $4.25\pm0.03$ & $15.51\pm0.04$ & $44.87\pm0.03$ \\
                  & 0.18 & $4.38\pm0.04$ & $15.43\pm0.05$ & $45.23\pm0.07$ \\
\hline
\multirow{2}{*}{17.6} & 0 & $4.47\pm0.03$ & $15.14\pm0.05$ & $45.03\pm0.04$ \\
                    & 0.1 & $4.60\pm0.04$ & $15.08\pm0.06$ & $45.40\pm0.07$ \\
\hline
\multirow{3}{*}{36.2} & 0 & $4.44\pm0.02$ & $15.12\pm0.03$ & $44.84\pm0.03$ \\
                   & 0.05 & $4.49\pm0.02$ & $15.09\pm0.03$ & $45.00\pm0.03$ \\
                    & 0.1 & $4.54\pm0.02$ & $15.07\pm0.03$ & $45.16\pm0.04$ \\
\enddata
\end{deluxetable}

Next, we assume that the line-of-sight absorption from X-ray and UV/optical are correlated, using the calibration of $N_{\rm H, host} = 5.55 \times 10^{21} \times E_{B-V, {\rm host}}$ \citep{Predehl1995}. The joint X-ray spectral analysis (\S\ref{subsec:xray_spec}) shows that $N_{\rm H, host}\approx 10^{21}\,{\rm cm^{-2}}$ at $t_{\rm obs}=7.8$\,d and decreases by a factor of 2--4 in the next two epochs. 
This implies an extinction from $E_{B-V, {\rm host}}\approx 0.18$ to $0.05 \lesssim E_{B-V, {\rm host}} \lesssim 0.1$. The best-fit blackbody models are plotted in Figure~\ref{fig:blackbody_SED}. As is shown in Table~\ref{tab:model_pars}, taking host extinction into consideration renders $R_{\rm bb}$ lower by a multiplicative factor of $\sim 1.1$, and $T_{\rm bb}$ and $L_{\rm bb}$ greater by multiplicative factors of $\sim1.3$ and $\sim2.2$, respectively.

By fitting the UV/optical SED as a blackbody, we are agnostic of the nature of this thermal component, which might be generated either by energy dissipation in stellar debris self-crossing shocks\footnote{Recent simulations by \citet{Huang2023_first_day} showed that the radiative luminosity from self-crossing shocks is much less than $10^{45}$\,erg\,s$^{-1}$ due to the effect of adiabatic losses.} \citep{Piran2015, Jiang2016_self_crossing_shock, Huang2023_first_day} or by reprocessing in an optically thick wind \citep{Miller2015_disk_wind, Metzger2016, Dai2018, Lu2020, Thomsen2022_disk_wind}.
A peak blackbody luminosity of $L_{\rm bb}\approx 10^{45}$\,erg\,s$^{-1}$ is on the high end of the bolometric luminosity function of ZTF-selected non-jetted TDEs (see Fig.~14 of \citealt{Yao2023}). 
The majority of such UV/optically overluminous TDEs do not exhibit broad emission lines of hydrogen, helium, and nitrogen that are commonly observed in TDEs \citep{vanVelzen2020}. Instead, many of them belong to the ``TDE-featureless'' subclass \citep{Hammerstein2023, Yao2023}, characterized by blue and featureless continuum emission in rest-frame UV and optical bands.

\subsection{X-ray: Internal Energy Dissipation in the Jet} \label{subsec:xray_syn}

\subsubsection{A Jet Synchrotron Model}
We consider an emitting plasma that is moving towards the observer at a bulk Lorentz factor $\Gamma_{\rm j}$. 
The emitting plasma is at a characteristic distance $r$ from the center of ejection, where $r$ is measured in the lab frame.
In the comoving frame of the plasma, the size of the causally connected region is $r/\Gamma_{\rm j}$. 
Therefore, we will only consider the emission from a roughly-spherical comoving volume of $\sim (r/\Gamma_{\rm j})^3$ --- what is beyond this volume cannot be correctly captured by the one-zone model considered here. 
The transverse size of the emitting region is $r/\Gamma_{\rm j}$, defining the area of emission as $\pi r^2 / \Gamma_{\rm j}^2$, which is the same in the lab and comoving frames. 
The longitudinal size of the emitting region is $r/\Gamma_{\rm j}$ in the comoving frame, and $r/\Gamma_{\rm j}^2$ in the lab frame due to length contraction.

Hereafter, all primed quantities are measured in the comoving frame of the emitting plasma. 
We omit the prime for the electron Lorentz factors ($\gamma$ and those with subscripts).
We assume that the plasma consists of electrons characterized by a broken-power-law distribution of Lorentz factors:
\begin{align}
    \frac{{\rm d} n^\prime }{{\rm d} \gamma} = \frac{n_{\rm b}^\prime}{\gamma_{\rm b}} {\rm min} \left[ (\gamma/\gamma_{\rm b})^{-p_1}, (\gamma/\gamma_{\rm b})^{-p_2} \right] ,
\end{align}
where $\gamma_{\rm b}$ is the break Lorentz factor, $n_{\rm b}^\prime$ denotes the comoving number density of electrons with $\gamma \sim \gamma_{\rm b}$, and $p_1$, $p_2$ are two power-law indices ($p_2>p_1$). 
The power-law indices are directly given by the photon indices below and above the break: $\Gamma_i = (p_i+1)/2$ for $i=1$, 2. In the \nustar observing epochs, $\Gamma_1 \sim 1.5$ and $\Gamma_2 \sim 2$, indicating that $p_1 \sim 2$ and $p_2 \sim 3$.

The dynamical timescale $t^\prime_{\rm dyn} = r/(\Gamma_{\rm j}c)$ can be described by the light-crossing time of the causally connected region.
We denote the cooling timescale for particles with Lorentz factor $\gamma$ as $t_{\rm c}^\prime (\gamma)$.
If particles are accelerated to a power-law distribution with injection index $p$ above a minimum Lorentz factor $\gamma_{\rm m}$ on a dynamical timescale and at the same time they
undergo (synchrotron+inverse-Compton) cooling, then depending on whether $t_{\rm c}^\prime (\gamma_{\rm m})$ is longer or shorter than $t^\prime_{\rm dyn}$, we may have two possible cases \citep{Sari1998}
\begin{enumerate}
    \item[] \textbf{Case 1:} In the ``slow-cooling'' case, $t_{\rm c}^\prime (\gamma_{\rm m}) > t^\prime_{\rm dyn}$, $p_1 = p$, $p_2 = p+1$, and $\gamma_{\rm b} = \gamma_{\rm c}$. 
    \item[] \textbf{Case 2:} In the ``fast-cooling'' case, $t_{\rm c}^\prime (\gamma_{\rm m}) < t^\prime_{\rm dyn}$,  $p_1 = 2$, $p_2 = p+1$, and $\gamma_{\rm b} = \gamma_{\rm m}$. 
\end{enumerate}
Here the cooling Lorentz factor $\gamma_{\rm c}$ is defined such that $t_{\rm c}^\prime (\gamma_{\rm c}) \equiv t^\prime_{\rm dyn}$. The observed X-ray spectral shape of AT2022cmc may be consistent with either of the two cases as long as the injection index is $p\approx 2$.

The magnetic field strength in the plasma is denoted as $B^\prime$. 
An electron with a Lorentz factor $\gamma$ has a characteristic synchrotron frequency of
\begin{align}
    \nu^\prime = \frac{3\gamma^2 eB^\prime}{4\pi m_{\rm e} c}, \label{eq:nu_syn}
\end{align}
and the peak specific power is given by (see, e.g., \citealt{Ghisellini2013})
\begin{align}
    \left(P_{\nu_{\rm p}} \right)^\prime \simeq \frac{\sqrt{3} e^3 B^\prime}{m_{\rm e} c^2}.
\end{align}
Consequently, the synchrotron emissivity at the break frequency $\nu_{\rm b}^\prime$ is given by
\begin{align}
    j^{\prime}_{\nu^\prime_{\rm b}} \simeq \left. \left( \gamma \frac{{\rm d}n^\prime}{{\rm d} \gamma}\right) \right|_{\gamma = \gamma_{\rm b}}  \left(P_{\nu_{\rm p}} \right)^\prime 
    =\frac{\sqrt{3} n_{\rm b}^\prime e^3 B^\prime}{m_{\rm e} c^2}.
\end{align}

The intensity at the surface of the emitting plasma is given by $I^{\prime}_{\nu^\prime_{\rm b}} = j^{\prime}_{\nu^\prime_{\rm b}} (r/\Gamma_{\rm j})$, and the lab-frame intensity is $I_{\nu_{\rm b}} = \Gamma_{\rm j}^3 I^{\prime}_{\nu^\prime_{\rm b}}$. The observed flux density at a distance of $d$ (without considering cosmological effects) is:
\begin{align}
    F_{\nu_{\rm b}} \simeq \frac{I_{\nu_{\rm b}} \pi r^2 / \Gamma_{\rm j}^2 }{ d^2 }.
\end{align}
Therefore, the isotropic equivalent spectral luminosity at the break frequency $\nu_{\rm b} \approx \Gamma_{\rm j} \nu_{\rm b}^\prime$ is given by
\begin{align}
    L_{\rm b, iso}  \equiv %\nu_{\rm b} L_{\nu_{\rm b}}
    4\pi d^2 \nu_{\rm b} F_{\nu_{\rm b}} %\notag & 
    \simeq 4\pi^2 r^3 \nu_{\rm b} \frac{\sqrt{3} n_{\rm b}^\prime e^3 B^\prime}{m_{\rm e} c^2}. \label{eq:L_biso}
\end{align}
The radiation energy density at a radius $r$ from the center is 
\begin{align}
    U^\prime_{\rm X} &= \frac{L_{\rm b, iso}}{4\pi r^2 \Gamma_{\rm j}^2 c} \notag \\
    &= 106\,{\rm erg\,cm^{-3}} L_{\rm b, iso, 47} r_{15}^{-2} (\Gamma_{\rm j}/50)^{-2},
\end{align}
and the magnetic energy density is $U^\prime_{\rm B} = B^{\prime 2} / (8\pi)$.

The break frequency is (cf. Eq.~\ref{eq:nu_syn})
\begin{align}
    \nu_{\rm b} = \frac{3 \Gamma_{\rm j} \gamma_{\rm b}^2 eB^\prime}{4\pi m_{\rm e} c} \label{eq:nu_b}.
\end{align}
To take into account cosmological effects, we use $\nu_{\rm b} = \nu_{\rm b, obs}(1+z)$ and $L_{\rm b, iso} = 4\pi d_L^2 \nu_{\rm b, obs} F_{\rm b, obs}$, where $\nu_{\rm b, obs}$ is the observed break frequency, and $F_{\rm b, obs}$ is the measured flux density at the observed break frequency. 

Suppose a certain mechanism puts a fraction $\epsilon_{\rm e}$ of the local energy into non-thermal electrons and another fraction $\epsilon_{\rm B}$ of the energy into magnetic fields, we see that $\epsilon_{\rm e} / \epsilon_{\rm B} \simeq U^\prime_{\rm X} / U^{\prime}_{\rm B}$, because an order unity fraction of the electrons' energy must be radiated in the X-ray band in either Case 1 or 2. For the same energy dissipation mechanism, we may expect the ratio $\epsilon_{\rm e} / \epsilon_{\rm B} $ to be constant in different epochs, which motivates us to define a single variable 
\begin{align}
    \xi_{\rm eB} \equiv \frac{\epsilon_{\rm e}}{\epsilon_{\rm B}} = \frac{U^\prime_{\rm X} }{U^{\prime}_{\rm B}} = \frac{2 L_{\rm b, iso}}{r^2 \Gamma_{\rm j}^2 c B^{\prime 2}}. \label{eq:xi_eB}
\end{align}

If we treat $\xi_{\rm eB}$ as a known, then we have three equations (\ref{eq:L_biso}, \ref{eq:nu_b}, and \ref{eq:xi_eB}) for five unknowns ($\Gamma_{\rm j}$, $r$, $n_{\rm b}^\prime$, $B^\prime$, $\gamma_{\rm b}$). We can then express three of the unknowns ($r$, $n_{\rm b}^\prime$, $B^\prime$) as functions of two independent variables ($\Gamma_{\rm j}$, $\gamma_{\rm b}$):
\begin{align}
    B^\prime & = \frac{4\pi m_{\rm e} c \nu_{\rm b}}{3\Gamma_{\rm j} \gamma_{\rm b}^2 e} , \\
    r &= \left( \frac{2L_{\rm b, iso}}{\xi_{\rm eB} c}\right)^{1/2} \frac{1}{\Gamma_{\rm j} B^\prime} , \\ 
    n_{\rm b}^\prime &= \frac{ L_{\rm b, iso} m_{\rm e} c^2 }{ 4\sqrt{3} \pi^2 r^3 \nu_{\rm b} e^3 B^\prime }.
\end{align}

% The key question is the physical identification of the break frequency so as to decide between the fast- and slow- cooling cases. 
Ignoring inverse-Compton cooling which is strongly Klein-Nishina suppressed (to be justified later), we consider the synchrotron cooling timescale
\begin{align}\label{eq:synchrotron_cooling_timescale}
    t_{\rm c}^\prime (\gamma_{\rm b}) = \frac{6\pi m_{\rm e}c}{\sigma_{\rm T} \gamma_{\rm b} B^{\prime 2} }.
\end{align}
The cooling frequency is $\nu_{\rm c} = \Gamma_{\rm j} \nu^\prime (\gamma_{\rm c})$.

\subsubsection{Solutions under Additional Constraints}

\begin{figure*}[htbp!]
    \centering
    \includegraphics[width=0.32\textwidth]{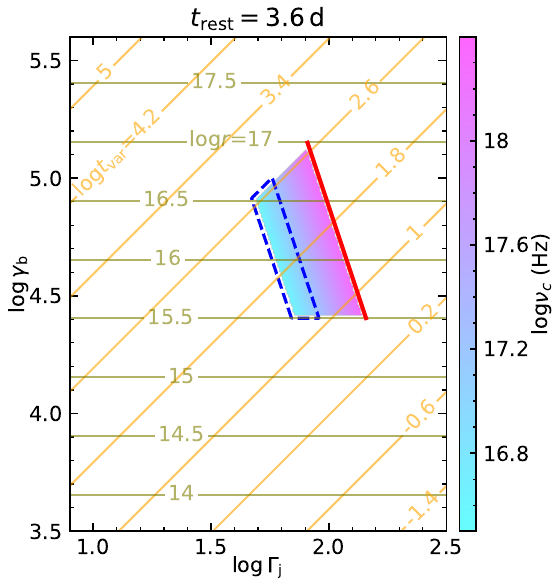}
    \includegraphics[width=0.32\textwidth]{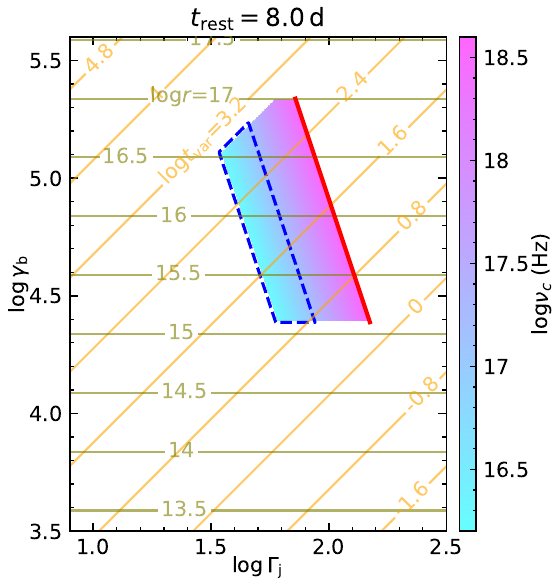}
    \includegraphics[width=0.32\textwidth]{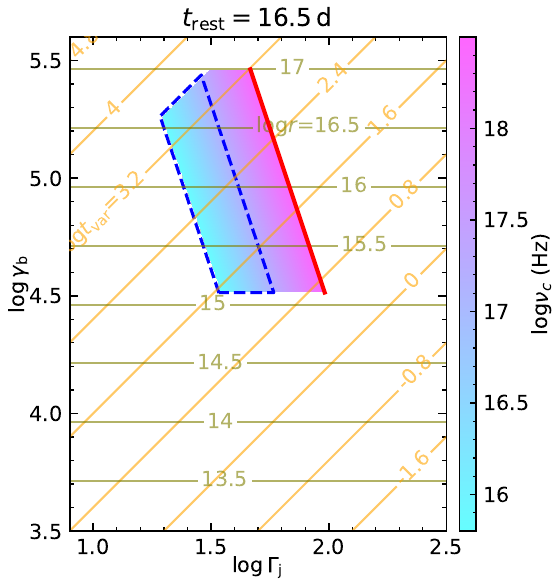}
    \caption{The colormaps show the parameter space for three epochs of \nustar observations that satisfy $\nu_{\rm c, min}<\nu_{\rm c} \leq \nu_{\rm b}$, $r<10^{17}$\,cm, $r>r_{\rm opt}$, and the limits on $t_{\rm var}$. $\xi_{\rm eB}=1$ is assumed.
    The orange lines show constant values of log($t_{\rm var}$/s), and the olive lines show constant values of log($r$/cm).
    The solid red lines along the upper right edge of the colormaps mark the parameter space in the slow cooling solution, where $\nu_{\rm c} = \nu_{\rm b}$. 
    The dashed blue lines mark the parameter space in the fast cooling solution, where $\nu_{\rm m} = \nu_{\rm b}$.
    \label{fig:xray_syn_pars}}
\end{figure*}

The system of equations is subjected to some additional constraints. 
First, we have an upper limit for the X-ray variability timescale $t_{\rm var} = t^\prime_{\rm dyn} / \Gamma_{\rm j}$. 
Second, there is a lower limit to the cooling frequency ($\nu_{\rm c}> \nu_{\rm c, min}$) so as to avoid the low-frequency tail of the synchrotron emission overproducing the thermal optical emission (see the dotted lines in Figure~\ref{fig:22cmc_sed}).
Third, in either the slow- or fast-cooling cases, the cooling frequency must not exceed $\nu_{\rm b}$. 
Next, the X-ray emitting radius must not exceed the emitting radius of the radio afterglow, which is constrained to be of the order $10^{17}$\,cm from \S\ref{subsec:radio_syn}, so we require $r<10^{17}$\,cm.
Finally, the X-ray emitting radius should be outside of the photosphere of the UV/optical thermal envelope (to be justified later).

\begin{deluxetable*}{c|c|ccc|ccc}[htbp!]
\tablecaption{Constraints on $\Gamma_{\rm j}$ and $\gamma_{\rm m}$ using the jet synchrotron model. \label{tab:gammas}}
\tablehead{
\colhead{$\xi_{\rm eB}$} 
& \colhead{} 
& \colhead{}
& \colhead{Case 1} 
& \colhead{}
& \colhead{}
& \colhead{Case 2}
& \colhead{}\\
\colhead{} 
& \colhead{} 
& \colhead{3.6\,d} 
& \colhead{8.0\,d} 
& \colhead{16.5\,d} 
& \colhead{3.6\,d} 
& \colhead{8.0\,d} 
& \colhead{16.5\,d} 
}
\startdata
\multirow{2}{*}{1} & $\Gamma_{\rm j}$ ($10^1$) & 8--14 & 7--15 & 5--10 & 5--9 & 3--9 & 2--6\\
                   & $\gamma_{\rm m}$ ($10^3$) & 3--36 & 1.5--43 & 1.5--65 & 25--100 & 24--171 & 33-274\\
\hline
\multirow{2}{*}{10} & $\Gamma_{\rm j}$ ($10^1$) & 5--8 & 4--8 & 3--5 & 3--5 & 2--5 & 1--3 \\
                    & $\gamma_{\rm m}$ ($10^3$) & 5--41 & 3--59 & 3--89 & 45--116 & 43--198 & 58--315\\
\hline
\multirow{2}{*}{0.1} & $\Gamma_{\rm j}$ ($10^1$) & 14--26 & 13--27 & 8--17 & 7--16 & 5--16 &3--10 \\
                    & $\gamma_{\rm m}$ ($10^3$) & 2--20 & 0.9--24 &  0.8--36 & 14--80 & 14--122 & 18--163\\
\enddata
\tablecomments{Case 1 is slow cooling; Case 2 is fast cooling.}
\end{deluxetable*}

In Figure~\ref{fig:xray_syn_pars}, we show the remaining parameter space under the above constraints, assuming $\xi_{\rm eB}=1$. 
The results depend weakly on $\xi_{\rm eB}$\footnote{As shown in Table~\ref{tab:gammas}, the exact constraints on $\Gamma_{\rm j}$ and $\gamma_{\rm m}$ varies by a factor of $<2$ when $\xi_{\rm eB}$ varies by a factor of 10.}.
For the first \nustar epoch (\S\ref{subsubsec:joint1}), we take $L_{\rm b, iso}=1.3\times 10^{47}$\,erg\,s$^{-1}$, $\nu_{\rm b}=10^{18.4}$\,Hz, $\nu_{\rm c, min} = 10^{16.5}$\,Hz, $t_{\rm var}\lesssim 500$\,s, and $r_{\rm opt}\approx 10^{15.5}$\,cm. 
For the second epoch (\S\ref{subsubsec:joint2}), we take $L_{\rm b, iso}=6\times 10^{46}$\,erg\,s$^{-1}$, $\nu_{\rm b}=10^{18.6}$\,Hz, $\nu_{\rm c, min} = 10^{16.2}$\,Hz, $t_{\rm var}\lesssim 10^3$\,s, and $r_{\rm opt}\approx 10^{15.1}$\,cm. 
For the third epoch, we assume that $\nu_{\rm b}$ still lies in the X-ray band and adopt the broken power-law fit in \S\ref{subsubsec:joint3}. We take $L_{\rm b, iso}=1.2\times 10^{46}$\,erg\,s$^{-1}$, $\nu_{\rm b}=10^{18.5}$\,Hz, $\nu_{\rm c, min} = 10^{15.8}$\,Hz, $t_{\rm var}\lesssim 1$\,hr, and $r_{\rm opt}\approx 10^{15.1}$\,cm. 
We see that the bulk Lorentz factor of the jet is relatively high in the allowed parameter space. Specifically, we have
\begin{align}
    10 \lesssim \Gamma_{\rm j} \lesssim 100.
\end{align}

A consequence of the high bulk Lorentz factor is that, if the X-ray emission occurs below the optical photospheric radius $r_{\rm opt}\sim 10^{15}$\,cm (\S\ref{subsec:uvopt_bbody}), the external radiation field will contribute a rather high energy density in the jet's comoving frame
\begin{align}
    U_{\rm opt}^\prime &\simeq U_{\rm opt} \Gamma_{\rm j}^2 = \frac{\tau_{\rm opt} \Gamma_{\rm j}^2 L_{\rm opt}}{ 4\pi r^2 c}\notag \\
    &\simeq 6.6\times 10^6 {\rm erg\, cm^{-3}} \tau_{\rm opt} (\Gamma_{\rm j}/50)^2 r_{15}^{-2} L_{\rm opt, 45},
\end{align}
where $\tau_{\rm opt}$ is the Rosseland-mean optical depth of the gas that is responsible for the thermal optical emission and $L_{\rm opt}\sim 10^{45}$\,erg\,s$^{-1}$ is the optical luminosity. 
Notably, the ratio $ U^\prime_{\rm opt} / U^\prime_{\rm X} = \tau_{\rm opt} \Gamma_{\rm j}^4 L_{\rm opt} /L_{\rm b, iso} \gg 1$. Such a high radiation energy density would lead to a very short cooling timescale: %\todo{double check with Wenbin}
\begin{align}\label{eq:external_inverse_Compton_cooling_timescale}
    t_{\rm c}^\prime (\gamma) & \simeq \frac{ \gamma m_{\rm e}c^2 }{ \gamma^2 U^\prime_{\rm opt} \sigma_{\rm T} c } \frac{1}{ Y_{\rm KN}(\gamma ) } \notag \\
    &= \frac{ m_{\rm e}c }{ \gamma U^\prime_{\rm opt} \sigma_{\rm T}  } \frac{ 1 }{ {\rm min} \left[ 1, \left( \frac{ m_{\rm e} c^2  }{\gamma \mathcal{E}_{\rm opt}^\prime } \right) ^2\right]  }\notag \\
    & = \frac{(6.2\times 10^{-4}\,{\rm s})  \; r_{15}^2}{\tau_{\rm opt} \gamma_4 (\Gamma_{\rm j}/50)^2 L_{\rm opt, 45}}{\rm max} \left[ 1, 9.8 \gamma_4 \frac{\Gamma_{\rm j}}{50} \frac{\mathcal{E}_{\rm opt}}{10\,{\rm eV}}\right]^2,
\end{align}
where the Klein-Nishina effects have been included here via the correction factor $Y_{\rm KN}\leq 1$, and $\mathcal{E}^\prime_{\rm opt} = \Gamma_{\rm j} \mathcal{E}_{\rm opt} \simeq 10\Gamma_{\rm j}$\,eV is the average photon energy in the comoving frame. 
By comparing the cooling time to the dynamical time $t^\prime_{\rm dyn} = r/(\Gamma_{\rm j}c)=667\,{\rm s}\,r_{15}(\Gamma_{\rm j}/50)^{-1}$, it becomes evident that at small radii $r \lesssim r_{\rm opt} \sim 10^{15}$\,cm, the intense radiation field of optical photons will rapidly cool relativistic electrons in the jet down to $\gamma \approx 1$. Thus, these low-Lorentz-factor electrons on their cooling track to $\gamma\approx 1$ will unavoidably overproduce the optical emission because $\nu_{\rm c,min}$ will be below the optical band (cf. Figure~\ref{fig:22cmc_sed}).
% Practically no relativistic power-law electron population can exist in the jet below $r_{\rm opt}$.

Alternatively, if the X-ray emission is generated above the optical photosphere, %i.e., $r\gg 10^{15}$\,cm, 
the optical photons will move nearly parallel to the jet's motion. According to Lorentz transformation, this leads to a considerable reduction in the energy density $U^\prime_{\rm opt}$ in the jet's comoving frame. 
As a result, a highly relativistic broken power-law electron population can exist. 
Therefore, a requirement of $r>r_{\rm opt}$ is necessary for the synchrotron model to remain viable. 
As can be seen in in Figure~\ref{fig:xray_syn_pars}, this constraint eliminates a substantial portion of the parameter space. 

Finally, we return to the question of the physical identification of the X-ray break frequency $\nu_{\rm b}$.

In the slow-cooling case, we have $\nu_{\rm c} = \nu_{\rm b}> \nu_{\rm m}$ and the solutions lie on the upper right edge of the surviving parameter space shown as red solid lines in Figure~\ref{fig:xray_syn_pars}. 
In different epochs, we find $\Gamma_{\rm j}\sim 100$ and $\gamma_{\rm c}=\gamma_{\rm b}\sim 10^{5}$. 
In this case, we require $\nu_{\rm m}>\nu_{\rm c, min}$ so as to avoid overproducing the UV/optical emission.
We also require $\nu_{\rm m}<10^{17.2}$\,Hz which corresponds to observer-frame 0.3\,keV, otherwise we should have observed two energy breaks in the X-ray band. 
This means that $0.1 \lesssim \gamma_{\rm m}/\gamma_{\rm b}\lesssim 0.2$ or $ \gamma_{\rm m}\sim 10^4$. 
The exact constraints on $\Gamma_{\rm j}$ and $\gamma_{\rm m}$ are presented in Table~\ref{tab:gammas}.

If the electrons are accelerated by internal shocks within a baryonic jet, $\gamma_{\rm m}\sim m_{\rm p}/ m_{\rm e} \Gamma_{\rm rel} \epsilon_e $, where $\Gamma_{\rm rel}$ is the relative Lorentz factor between two colliding shells. When $\Gamma_{\rm rel}\gtrsim 10$, acceleration by internal shocks is possible. 
A magnetically dominated jet with high magnetization is also plausible, as long as the magnetic energy in the average volume per electron is as high as $\gamma_{\rm m} m_{\rm e} c^2$. For instance, if the jet is made of proton-electron plasma and particles are initially cold in the comoving frame, then the magnetization is given by $\sigma \simeq B^{\prime 2} / (8\pi)/(n^\prime m_{\rm p} c^2)$, and as long as $\sigma \sim \gamma_{\rm m} m_{\rm e}/m_{\rm p}\gtrsim 10$, our solution is plausible.

In the fast-cooling case, we have $\nu_{\rm m} = \nu_{\rm b} > \nu_{\rm c}$. Since only one energy break was observed in the X-ray band, we have $\nu_{\rm c, min}< \nu_{\rm c}<10^{17.2}$\,Hz, and the solutions lie within the regions marked by the dashed blue lines in Figure~\ref{fig:xray_syn_pars}. In different epochs, we find the bulk Lorentz factor to be smaller than in the slow-cooling case, $20\lesssim \Gamma_{\rm j}\lesssim 90$, and the minimum Lorentz factor for particle acceleration to be greater, $\gamma_{\rm m} = \gamma_{\rm b}\sim 10^5$. 
Such a high $\gamma_{\rm m}$ disfavors internal shocks within a baryonic jet. A magnetically dominated jet with high magnetization $\sigma\sim100$ is a plausible solution.

Now that we have reasonable solutions for the Lorentz factor of the electrons radiating near the break frequency $\gamma_{\rm b} \sim 10^5$, we go back to the question whether their cooling is dominated by synchrotron (as assumed in Eq. \ref{eq:synchrotron_cooling_timescale}) or synchrotron-self-Compton. The typical energy of the X-ray photons in the electron's rest frame is
\begin{equation}
    \mathcal{E}_{\rm X}'\sim {\gamma_{\rm b}\over \Gamma_{\rm j}} \mathcal{E}_{\rm X} \simeq 20\mathrm{\,MeV}\, {\gamma_{\rm b}/10^5 \over \Gamma_{\rm j}/50} {\mathcal{E}_{\rm X}\over 10\mathrm{\, keV}}.
\end{equation}
This leads to a large Klein-Nishina suppression factor of $Y_{\rm KN}\sim (m_{\rm e} c^2/\mathcal{E}_{\rm X}')^2 \sim 10^{-3}$ (for fiducial parameters in the above expression) for the inverse-Compton scattering power as compared to the case of Thomson scattering. This means that synchrotron-self-Compton cooling rate is a factor of $Y_{\rm KN}\xi_{\rm eB}\ll 1$ smaller than synchrotron cooling rate, so Eq. (\ref{eq:synchrotron_cooling_timescale}) is justified.

Unfortunately, current observations cannot break the degeneracy between the two cases. Additional knowledge about the particle number density (or magnetization) in the jet is needed --- this piece of information may be obtained if the source is at a closer distance such that we detect the synchrotron self-Compton emission in the high-energy gamma-ray band.

%\begin{figure*}[htbp!]
%    \centering
%    \includegraphics[width=0.32\textwidth]{para_scan_1_xieB_10.0.pdf}
%    \includegraphics[width=0.32\textwidth]{para_scan_2_xieB_10.0.pdf}
%    \includegraphics[width=0.32\textwidth]{para_scan_3_xieB_10.0.pdf}
%    \caption{Same as in Figure~\ref{fig:xray_syn_pars}, but for $\xi_{\rm eB} = 10$.  \label{fig:xray_syn_pars_xieB10}}
%\end{figure*}

%\begin{figure*}[htbp!]
%    \centering
%    \includegraphics[width=0.32\textwidth]{para_scan_1_xieB_0.1.pdf}
%    \includegraphics[width=0.32\textwidth]{para_scan_2_xieB_0.1.pdf}
%    \includegraphics[width=0.32\textwidth]{para_scan_3_xieB_0.1.pdf}
%    \caption{Same as in Figure~\ref{fig:xray_syn_pars}, but for $\xi_{\rm eB} = 0.1$.  \label{fig:xray_syn_pars_xieB10}}
%\end{figure*}

\section{Discussion} \label{sec:discuss}

Using \nicer, \swift, and \xmm, we constructed the 0.3--10\,keV X-ray light curve of the on-axis jetted TDE AT2022cmc, which roughly follows $L_{\rm X}\propto t^{-2}$ (Figure~\ref{fig:xlc}) in the first $\sim10$\,months of evolution.
At late time during the X-ray evolution of other on-axis jetted TDEs, a sudden flux drop has been observed in both Sw\,J1644+57 (at rest-frame days since discovery $t_{\rm rest} \approx 370$\,days; \citealt{Zauderer2013}) and Sw\,J2058+05 ($t_{\rm rest} \approx 200$\,days; \citealt{Pasham2015}), which has been explained by a jet shut-off as the accretion flow transitions from a supercritical thick disk to a geometrically thin disk state \citep{Tchekhovskoy2014}. Such a disk instability triggered state transition is naturally predicted as a result of the decreasing mass accretion rate \citep{Shen2014, Lu2022}, and has recently also been observed in the non-jetted TDE AT2021ehb \citep{Yao2022}. Future \chandra X-ray monitoring observations of AT2022cmc will reveal if the luminosity continues to follow the power-law decay and verify the existence of such a disk state transition. 

AT2022cmc is the first on-axis jetted TDE ever observed with \nustar. 
Joint X-ray spectral analysis between \nustar and \nicer reveals a broken powerlaw spectral shape. 
We interpret the X-rays as synchrotron emission generated by internal dissipation within the jet. 
The inferred jet Lorentz factor $\Gamma_{\rm j}\sim 50$, which is higher than the $\Gamma_{\rm j}\approx 10$ estimated in Sw\,J1644+57 \citep{Bloom2011}. Unfortunately, our current model does not constrain the physical opening angle of the jet, which may be wider than $1/\Gamma_{\rm j}\simeq 1^{\circ} (\Gamma_{\rm j}/50)$. For instance, the jet opening angles of GRBs are typically much wider than $1/\Gamma_{\rm j}$ for their high Lorentz factors $\Gamma_{\rm j}\gtrsim 100$ \citep{Frail2001}. However, jets in jetted TDEs must be strongly beamed because of their low rates.

%%% rate
The rate of on-axis jetted TDEs with prompt X-ray luminosity above $10^{48}$\,erg\,s$^{-1}$ has been estimated to be $0.03_{-0.02}^{+0.04}$\,Gpc$^{-3}$\,yr$^{-1}$ from \swift/BAT \citep{Sun2015}.
The volumetric rates of TDEs with soft X-ray and UV/optical thermal emission above $10^{43}$\,erg\,s$^{-1}$ is $\sim\!230$\,Gpc$^{-3}$\,yr$^{-1}$ and $\sim\!310$\,Gpc$^{-3}$\,yr$^{-1}$, respectively \citep{Sazonov2021, Yao2023}. Given the recent infrared discovery of a sample of mid-IR TDE candidates in nearby galaxies \citep{jiang21_MIRONG} and another nearby heavily dust-extincted TDE \citep{Panagiotou2023}, we assume that a comparable fraction of TDEs are missed by soft X-ray/optical time domain surveys, implying a total TDE rate of the order $\sim\! 10^3$\,Gpc$^{-3}$\,yr$^{-1}$. This means that only a very small fraction, $\sim\! 3\times10^{-5}$, of TDEs have bright, on-axis jet X-ray emission. Part of the reason for this small fraction is the beaming factor of jets $f_{\rm b}> \pi \Gamma_{\rm j}^{-2}/2=6\times10^{-4}(\Gamma_{\rm j}/50)^{-2}$, but we see that beaming alone does not account for the low detection rate of jetted TDEs. In fact, less than a few percent of TDEs are intrinsically jetted --- possibly due to the lack of a strong magnetic flux \citep{Tchekhovskoy2014, kelley14_magnetic_flux} or slow black hole spins \citep{Andreoni2022} in most cases, or a large fraction of jets are hydrodynamically choked by the surrounding gas \citep{DeColle12_jet_dynamics}.

% This implies that the fraction of TDEs that launch relativistic jets is $\sim10$\%.

%\begin{acknowledgements}

\vspace{1cm}

\textit{Acknowledgements} -- 
We thank the anonymous reviewer for providing constructive comments.
YY thanks Dheeraj Pasham and Keith Gendreau for sharing quicklook \nicer data. She also thanks Eli Waxman, Bing Zhang, Matteo Lucchini and Igor Andreoni for helpful discussions. 

%%% NASA GRANT acknowledgement (NuSTAR Cycle 7; NICER Cycle 4)
YY acknowledges support from NASA under award No. 80NSSC22K0574 and No. 80NSSC22K1347.

%%% NuSTAR standard acknowledgement
This work has made use of data from the \nustar mission, a project led by Caltech, managed by NASA/JPL, and funded by NASA. 
This research has made use of the NuSTAR Data Analysis Software (NuSTARDAS), jointly developed by the ASI Science Data Center (ASDC, Italy) and the Caltech (USA).

%\end{acknowledgements}

%\appendix

\bibliography{main}{}
\bibliographystyle{aasjournal}

\end{document}